\documentclass{emulateapj}
\newcommand{\chandra}{{\it Chandra}}
\newcommand{\bima}{{BIMA}}
\newcommand{\ovro}{{OVRO}}
\newcommand{\sza}{{SZA}}
\newcommand{\rtfh}{\mbox{$r_{\mbox{\scriptsize 2500}}$}}

\newcommand{\rfh}{\mbox{$r_{\mbox{\scriptsize 500}}$}}

\newcommand{\Mtfh}{\mbox{$M_{\mbox{\scriptsize 2500}}$}}

\newcommand{\Mfh}{\mbox{$M_{\mbox{\scriptsize 500}}$}}

\newcommand{\Ys}{\mbox{$Y_{\mathrm{sph}}$}}
\newcommand{\Mwl}{\mbox{$M_{\mathrm{WL}}$}}
\newcommand{\aYM}{\mbox{$B$}}
\newcommand{\SigYM}{\mbox{$\sigma_{M|Y}$}}

\newcommand{\my}{\mbox{\Mwl$-$\Ys}}
\newcommand{\Msun}{M$_\odot$}
\newcommand{\bq}{\begin{equation}}
\newcommand{\eq}{\end{equation}}
\newcommand{\oke}{\citetalias{OkabeE10}}
\newcommand{\ww}{\mbox{$\left<w\right>$}}
\newcommand{\ergs}{\mbox{${\rm erg\,s^{-1}}$}}

\def\ls{\mathrel{\hbox{\rlap{\hbox{\lower4pt\hbox{$\sim$}}}\hbox{$<$}}}}
\def\gs{\mathrel{\hbox{\rlap{\hbox{\lower4pt\hbox{$\sim$}}}\hbox{$>$}}}}

\defcitealias{SmithE05}{Sm05}
\defcitealias{OkabeE10}{Ok10}

\shorttitle{LoCuSS: SZ Effect and Weak-lensing Scaling Relation}
\shortauthors{Marrone et al.}
\slugcomment{The Astrophysical Journal, 754:119, 2012 August 1}

\begin{document}

\title{LoCuSS: The Sunyaev-Zel'dovich Effect and Weak Lensing Mass Scaling Relation}

\author{
Daniel P.\ Marrone,$\!$\altaffilmark{1,2,20}
Graham P.\ Smith,$\!$\altaffilmark{3}
Nobuhiro Okabe,$\!$$\!$\altaffilmark{4,5}
Massimiliano~Bonamente,$\!$\altaffilmark{6,7} 
John~E.~Carlstrom,$\!$\altaffilmark{1,8,9,10}
Thomas~L.~Culverhouse,$\!$\altaffilmark{11,12}
Megan~Gralla,$\!$\altaffilmark{1,8}
Christopher~H.~Greer,$\!$\altaffilmark{1,8}
Nicole~Hasler,$\!$\altaffilmark{6} 
David~Hawkins,$\!$\altaffilmark{12}
Ryan~Hennessy,$\!$\altaffilmark{1,8}
Marshall~Joy,$\!$\altaffilmark{7}
James~W.~Lamb,$\!$\altaffilmark{12}
Erik~M.~Leitch,$\!$\altaffilmark{1,8}
Rossella Martino,$\!$\altaffilmark{13}
Pasquale Mazzotta,$\!$\altaffilmark{13}
Amber~Miller,$\!$\altaffilmark{14,15}
Tony~Mroczkowski,$\!$\altaffilmark{16,21}
Stephen~Muchovej,$\!$\altaffilmark{12}
Thomas~Plagge,$\!$\altaffilmark{1,8}
Clem~Pryke,$\!$\altaffilmark{17}
Alastair J.\ R.\ Sanderson,$\!$\altaffilmark{3}
Masahiro Takada,$\!$$\!$\altaffilmark{18}
David~Woody$\!$\altaffilmark{12}
and Yuying~Zhang$\!$\altaffilmark{19}
}
\altaffiltext{1}{Kavli Institute for Cosmological Physics, University of Chicago, Chicago, IL}
\altaffiltext{2}{Steward Observatory, University of Arizona, Tucson, AZ}
\altaffiltext{3}{School of Physics and Astronomy, University of Birmingham, Edgbaston, Birmingham, UK}
\altaffiltext{4}{Astronomical Institute, Tohoku University, Aramaki, Aoba-ku, Sendai, Japan}
\altaffiltext{5}{Institute of Astronomy and Astrophysics, Academia Sinica, Taipei, Taiwan, R.O.C.}
\altaffiltext{6}{Department of Physics, University of Alabama, Huntsville, AL}
\altaffiltext{7}{Space Science-VP62, NASA Marshall Space Flight Center, Huntsville, AL}
\altaffiltext{8}{Department of Astronomy and Astrophysics, University of Chicago, Chicago, IL}
\altaffiltext{9}{Enrico Fermi Institute, University of Chicago, Chicago, IL}
\altaffiltext{10}{Department of Physics, University of Chicago, Chicago, IL}
\altaffiltext{11}{Radio Astronomy Lab, University of California, Berkeley, CA}
\altaffiltext{12}{Owens Valley Radio Observatory, California Institute of Technology, Big Pine, CA USA}
\altaffiltext{13}{Dipartimento di Fisica, Universit\`a degli Studi di Roma ``Tor Vergata'', via della Ricerca Scientifica 1, Roma, Italy}
\altaffiltext{14}{Columbia Astrophysics Laboratory, Columbia University, New York, NY}
\altaffiltext{15}{Department of Physics, Columbia University, New York, NY}
\altaffiltext{16}{Department of Physics and Astronomy, University of Pennsylvania, Philadelphia, PA}
\altaffiltext{17}{School of Physics and Astronomy, University of Minnesota, Minneapolis, MN}
\altaffiltext{18}{Institute for the Physics and Mathematics of the Universe (IPMU), University of Tokyo, Chiba, Japan}
\altaffiltext{19}{Argelander-Institut f\"ur Astronomie, Universit\"at Bonn, Bonn, Germany}
\altaffiltext{20}{Hubble Fellow}
\altaffiltext{21}{Einstein Postdoctoral Fellow}

\email{dmarrone@email.arizona.edu}

\begin{abstract}
  We present the first weak-lensing-based scaling relation between
  galaxy cluster mass, \Mwl, and integrated Compton parameter \Ys.
  Observations of 18 galaxy clusters at $z\simeq0.2$ were obtained
  with the Subaru 8.2-m telescope and the Sunyaev-Zel'dovich Array.
  The \my\ scaling relations, measured at $\Delta$~=~500, 1000, and
  2500~$\rho_c$, are consistent in slope and normalization with
  previous results derived under the assumption of hydrostatic
  equilibrium (HSE).  We find an intrinsic scatter in \Mwl\ at fixed
  \Ys\ of 20\%, larger than both previous measurements of
  $M_\mathrm{HSE}-\Ys$ scatter as well as the scatter in true mass at
  fixed \Ys\ found in simulations.  Moreover, the scatter in our
  lensing-based scaling relations is morphology dependent, with 30 -- 40\%
  larger \Mwl\ for undisturbed compared to disturbed clusters at the
  same \Ys\ at \rfh. Further examination suggests that the segregation
  may be explained by the inability of our spherical lens models to
  faithfully describe the three-dimensional structure of the clusters,
  in particular, the structure along the line-of-sight.  We find that
  the ellipticity of the brightest cluster galaxy, a proxy for halo
  orientation, correlates well with the offset in mass from the mean
  scaling relation, which supports this picture. This provides 
  empirical evidence that line-of-sight
  projection effects are an important systematic uncertainty in
  lensing-based scaling relations.
\end{abstract}


\section{Introduction}

Galaxy clusters are tracers of the highest peaks in the matter density
field, and as such, their abundance as a function of mass and redshift
depends strongly upon cosmology.  Along with data from a variety of
other techniques \citep{KomatsuE11,HickenE09,KesslerE09,PercivalE10},
galaxy cluster surveys have placed precise constraints on the
$\Lambda$CDM cosmological parameters
\citep{VikhlininE09-CCCP3,MantzE10}.  To
improve the constraints from cluster surveys, the scaling relationship
between survey observables and the total cluster mass must be
precisely determined.

Surveys at millimeter wavelengths can be used to identify large
numbers of clusters via the Sunyaev-Zel'dovich (SZ) effect, as has
been demonstrated by the South Pole Telescope (SPT;
\citealt{StaniszewskiE09,VanderlindeE10,WilliamsonE11}), the Atacama
Cosmology Telescope (ACT; \citealt{MarriageE10}), and the {\it Planck}
satellite \citep{planck-esz}.  The SZ effect is a spectral distortion
of the cosmic microwave background (CMB) resulting from the inverse Compton
scattering of CMB photons by the hot electrons of the intra-cluster
medium (ICM).  The magnitude of this distortion is determined by the
integral of the pressure along the line of sight through the cluster.
The volume integral of pressure is the total thermal energy content of
the ICM, which should directly trace the depth of the potential well
\citep{CHR02}, so a tight scaling between the SZ signal ($Y$, see
Section~\ref{sec:szmodel}) and mass is expected.

A low-scatter correlation is indeed found in simulations of galaxy
clusters \citep[e.g.,][]{daSilvaE04,MotlE05,Nagai06,StanekE10}, and is
suggested in the best comparisons of SZ observations and cluster mass
estimates available to date
\citep{BonamenteE08,PlaggeE10,AnderssonE11}.  However, the mass-$Y$
scaling relation remains poorly determined from these observations,
and the uncertainty in the relation is already limiting the
cosmological constraints derived from small SZ-selected samples
\citep{VanderlindeE10,SehgalE11}.  Moreover, previous comparisons of $Y$ with
mass have relied on mass estimates derived assuming hydrostatic
equilibrium (HSE).  This assumption is certain to result in at least
some bias, although the degree of bias is poorly known; observational
constraints suggest $\lesssim20$\% within radii of interest
\citep{MahdaviE08,ZhangE10}, though some simulations have found larger
biases \citep[e.g.,][]{CavaliereE11}.

One promising technique for improving our knowledge of the mass-$Y$
relation is weak gravitational lensing
\citep[e.g.,][]{DahleE02,HoekstraE07,OkabeE10}.  Unlike other
techniques, weak lensing directly probes the gravitational potential
of the cluster, providing a way to test for mass biases in other mass
estimators.  However, the weak lensing signal is sensitive to the
total mass projected along the line-of-sight, and thus likely suffers
projection-related errors in mass measurements 
\citep[e.g.,][]{MetzlerE01,Hoekstra01,Hoekstra03}.  Several authors have
calibrated X-ray observables against lensing mass estimates, typically
finding consistency with (but with larger scatter than) X-ray-only
scaling relations
\citep{SmithE05,BardeauE07,HoekstraE07,ZhangE07,ZhangE08,RykoffE08,OkabeE10b,HoekstraE11}.
This technique has not yet been used to calibrate the SZ observable
except in the cluster core \citep{MarroneE09}. An important next step
is to make such a comparison at the larger radii typically used in
cosmological studies with galaxy clusters.

The calibration of mass-observable relations is one of the key goals
of the Local Cluster Substructure Survey
(LoCuSS\footnote{\url{http://www.sr.bham.ac.uk/locuss}}), a
multi-wavelength survey of galaxy clusters at $0.15 < z < 0.3$
selected from the {\it ROSAT} All-sky Survey
\citep{EbelingE98,EbelingE00,BohringerE04}.  The large luminosity range
and lack of morphological selection
in the sample ensure a morphologically diverse population for calibration studies.
Weak lensing mass estimates are available for an initial sample of 29
LoCuSS clusters (\citealt{OkabeE10}; hereafter \citetalias{OkabeE10}).
In this paper, we combine these estimates with SZ data from the
Sunyaev-Zel'dovich Array (SZA) in order to examine the mass-$Y$
scaling relation and its scatter.  We also explore the possibility of
biases in the mass estimates related to cluster morphology, which our
data suggest is an important systematic.

Section~\ref{sec:observations} describes the sample of clusters as
well as the weak lensing and SZ observations.  In
Section~\ref{sec:analysis}, we discuss the analysis techniques used to
derive the weak lensing mass estimates and determine the SZ
observables.  Our scaling relation results are presented in
Section~\ref{sec:results} and our interpretation of the findings is
discussed in Section~\ref{sec:discussion}.  Our conclusions are
reviewed in Section~\ref{sec:conclusions}. We assume a flat
$\Lambda$CDM cosmology: $\Omega_\mathrm{M}=0.27$,
$\Omega_\Lambda=0.73$, $H_0=73\rm~km~s^{-1}~Mpc^{-1}$.

\section{Observations}
\label{sec:observations}

\subsection{Cluster Sample}\label{sec:sample}

The clusters considered in this work are drawn from the LoCuSS 
``high-$L_{\rm X}$'' sample discussed by \citetalias{OkabeE10} (G.~P. Smith, in prep.), which was selected from the \emph{ROSAT}
All-Sky Survey to have $0.15 < z < 0.3$, $-20^\circ\le\delta\le+60^\circ$,
and $n_H<7\times10^{20}{\rm cm}^{-2}$.  This sample is subject to a luminosity limit 
of $L_\mathrm{X}/E(z)^{2.7}>4.1\times10^{44}$~ergs, where 
$E(z)$ describes the redshift evolution of the Hubble parameter
($E(z)^2\equiv H(z)^2/H_0^2=\Omega_M(1+z)^3+\Omega_\Lambda$ for a flat cosmology).
Specifically, we consider 21 
clusters from the high-$L_{\rm X}$ sample (Table~\ref{t:obs}) for which high quality $V$- and $i'$-band 
Subaru/Suprime-Cam data are available, and in which a central mass concentration is identifiable
in the shear maps \citepalias[see Table~5 of][]{OkabeE10}.
The sample spans a factor of $4$ in
X-ray luminosity and includes clusters with a broad range of X-ray
morphologies, including cool-core (hereafter CC) clusters
\citep[e.g.,\ A\,383, A\,1835:][]{SmithE01,SchmidtE01}, cold front
clusters \citep[e.g.,\ RX\,J1720.1$+$2638:][]{MazzottaE01}, and
clusters known to be undergoing mergers \citep[e.g.,\ A\,209:][and
references therein]{GiovanniniE09}.

\subsection{Gravitational Lensing}

The acquisition and processing of the
Subaru/Suprime-Cam\footnote{Based in part on data collected at Subaru
  Telescope and obtained from the SMOKA, which is operated by the
  Astronomy Data Center, National Astronomical Observatory of Japan.}
data upon which the weak lensing mass measurements are based are
described in detail by \citetalias{OkabeE10}.  Briefly, the data were
acquired in the $V$- and $i'$-bands in excellent atmospheric
conditions ${\rm FWHM}\simeq0.7''$, consistent with the stringent
requirements for precise measurement of galaxy shapes.  Weakly-lensed
background galaxies were selected to have a minimum color offset from
the cluster red sequence in the color-magnitude plane following
\citet{UmetsuBroadhurst08} and \citet{UmetsuE09}.  Background galaxy
shapes were measured using KSB \citep{KaiserE95}, achieving a residual
mean ellipticity of point sources after removal of the point spread
function of $\sim10^{-4}$, and a typical number density of galaxies
for use in the weak lensing analysis of $\sim10-30$~arcmin$^{-2}$.

\subsection{Sunyaev-Zel'dovich Effect}
\label{sec:sz}

The sample of 21 clusters was observed with the \sza, an 8-element
radio interferometer optimized for measurements of the SZ effect. The
SZA is a subset of the 23-element Combined Array for Research in
Millimeter-wave Astronomy (CARMA).  Data were obtained in an 8~GHz
passband centered at 31~GHz between 2006 and 2009 (Table~\ref{t:obs}).
Over the period of these observations, the array was sited at two
different locations and occupied three different configurations, as
described in \citet{CulverhouseE10}. For most of this period, the
\sza\ was configured with six elements in a compact array sensitive to
arcminute-scale SZ signals, and two outrigger elements providing
discrimination for compact radio source emission. The compact array
and outrigger baselines provide baseline lengths, in units of number
of wavelengths $\lambda$, of 0.35$-$1.3~k$\lambda$ and
2$-$7.5~k$\lambda$, respectively. Two of the clusters (Abell~209 and
ZwCl~1454.8+2233) were also briefly observed with an
eight-compact-element array configuration, with greater SZ sensitivity
but little intrinsic power to distinguish SZ signal from
contamination. Integration times were tailored to the magnitude of the
SZ signal in each cluster and, secondarily, to the degree of
contamination from radio sources. 
Clusters with complex radio source environments in 1.4~GHz survey images were given more integration time. 
The total on-source unflagged
integration times range from 5 to 69 hours, and the corresponding
noise levels achieved by the short (SZ-sensitive) baselines were 0.14
to 0.45~mJy (12 to 40~$\mu$K), with synthesized beams of
$\sim$2\arcmin\ FWHM. Absolute calibration of the data is determined
through periodic measurements of Mars, calibrated against the
\citet{RudyE87} model for this source.  A systematic uncertainty of
10\% applies to the SZ measurements due to the uncertainty in the
absolute calibration \citep{SharpE10}, this is not included in the reported errors.

Exclusion of emissive sources from the \sza\ data is most simply
achieved when the sources are spatially compact.  For such sources the
emission measured on the longer outrigger baselines (smaller angular
scales) can be assumed to be representative of the emission measured
on the short, SZ-sensitive baselines (larger angular scales) and
thereby separated from the SZ signal.  Sources that are not point-like
on the scale of 20\arcsec\ leave residual emission in the short
baseline maps that fills in the cluster SZ signal.  Three clusters in
the sample suffer from extended radio source contamination:
\begin{itemize}
\item Abell~115 hosts the bright radio source 3C~28, which is
  $>$100~mJy at 5~GHz and has lobes extending more than an arcminute
  across the cluster \citep{GiovanniniE87}.  While there is some
  discrimination on the short baselines between the extended central
  source and the more extended SZ decrement, reliable separation of
  these two resolved structures is not possible in these data.
\item RX~J1720.1+2638 contains a bright source near the cluster center
  that is resolved on scales of 30\arcsec\ at 1.4~GHz in the VLA FIRST
  survey \citep{FIRST}. The \sza\ data suggest that this source is
  also extended at 31~GHz, where the SZ decrement appears to be split
  through its center by an emissive source after a point-like source
  model is determined from the long baselines and removed.
\item Abell~291 hosts three 31~GHz sources within an arcminute of the
  pointing center.  One of these is revealed to be a double-lobed
  radio source in FIRST, and residual emission at this position
  appears to fill in the SZ decrement, leading to an offset of
  $>$20\arcsec\ between the SZ centroid and BCG position.
\end{itemize}
Observations of these objects in the 90~GHz SZA band would likely
provide greater spectral discrimination between the radio source
emission and SZ signal; this possibility will be examined through
future observations.  

In subsequent sections we analyze the sample of 18 clusters that remains after
excluding the three objects described above. As in \citetalias{OkabeE10}, we examine 
whether the 18-cluster subsample is representative of the parent sample by constructing 
randomized (with replacement) groups of the same size from the full sample. We find that the mean value of 
$L_\mathrm{X}/E(z)^{2.7}$ for our sample is very similar to the parent population, falling 
in the 47$^{th}$ percentile of random groups. The spread in $L_\mathrm{X}/E(z)^{2.7}$ within 
the sample is also very representative, the standard deviation in this quantity for our 
sample is again at the 47$^{th}$ percentile of the random groups.

\begin{deluxetable*}{lcccccccc}
\tabletypesize{\scriptsize}
\tablecolumns{8}
\tablewidth{0pt}
\tablecaption{Cluster Sample and SZ Observational Parameters\label{t:obs}}
\tablehead{
Cluster & R.A.\tablenotemark{a} & Dec\tablenotemark{a} & $z$ & $L_\mathrm{X}[0.1-2.4\mathrm{keV}]$ & $t_{int}$\tablenotemark{b} & $S_{rms}$\tablenotemark{c} & Resolution\tablenotemark{c,d} & $T_{rms}$\tablenotemark{c} \\
 & & & & ($10^{44}\ergs$) & (hrs) & (mJy) & (\arcsec) & ($\mu$K)}
 \startdata
Abell\,68            & 00:37:05.90 & $+$09:09:26.6 & 0.255 &  8.8 & 17.3 & 0.22 & 122$\times$139 & 17 \\
Abell\,115N	     & 00:55:50.38 & $+$26:24:36.0 & 0.197 &  8.6 &  7.7 & 0.36 & 126$\times$130 & 29 \\
Abell\,209           & 01:31:53.47 & $-$13:36:46.1 & 0.206 &  7.3 & 19.3 & 0.21 & 91$\times$125  & 24 \\
RX\,J0142.0$+$2131   & 01:42:02.64 & $+$21:31:19.2 & 0.280 &  5.9 & 11.6 & 0.30 & 114$\times$126 & 27 \\
Abell\,267           & 01:52:41.93 & $+$01:00:24.1 & 0.230 &  8.1 & 18.4 & 0.30 & 115$\times$160 & 21 \\
Abell\,291     	     & 02:01:43.11 & $-$02:11:48.1 & 0.196 &  5.7 & 27.8 & 0.22 & 101$\times$124 & 23 \\
Abell\,383           & 02:48:03.50 & $-$03:31:45.0 & 0.188 &  5.3 & 25.0 & 0.27 & 128$\times$159 & 18 \\
Abell\,521           & 04:54:06.90 & $-$10:13:24.6 & 0.247 &  9.5 & 12.8 & 0.32 & 117$\times$210 & 26 \\
Abell\,586           & 07:32:20.31 & $+$31:38:02.0 & 0.171 &  6.6 & 13.1 & 0.25 & 120$\times$137 & 21 \\
Abell\,611           & 08:00:56.74 & $+$36:03:21.6 & 0.288 &  8.1 & 30.2 & 0.25 & 120$\times$130 & 21 \\
Abell\,697           & 08:42:57.80 & $+$36:21:54.5 & 0.282 &  9.6 & 15.8 & 0.35 & 118$\times$128 & 30 \\
Abell\,1835          & 14:01:02.02 & $+$02:52:41.7 & 0.253 & 22.8 & 13.5 & 0.36 & 117$\times$149 & 27 \\
ZwCl\,1454.8+2233    & 14:57:15.09 & $+$22:20:34.2 & 0.258 &  7.8 & 68.6 & 0.14 & 119$\times$123 & 12 \\
ZwCl\,1459.4+4240    & 15:01:23.13 & $+$42:20:39.6 & 0.290 &  6.7 & 21.7 & 0.22 & 115$\times$131 & 19 \\
Abell\,2219          & 16:40:22.60 & $+$46:42:22.0 & 0.228 & 12.1 & 27.3 & 0.20 & 114$\times$133 & 17 \\
RX\,J1720.1+2638     & 17:20:09.90 & $+$26:37:27.8 & 0.164 &  9.5 & 26.9 & 0.19 & 108$\times$125 & 19 \\
Abell\,2261          & 17:22:27.08 & $+$32:07:58.6 & 0.224 & 10.8 &  9.9 & 0.31 & 114$\times$134 & 26 \\
RX\,J2129.6+0005     & 21:29:39.90 & $+$00:05:19.8 & 0.235 & 11.0 & 24.5 & 0.27 & 113$\times$167 & 19 \\
Abell\,2390          & 21:53:36.70 & $+$17:41:31.0 & 0.233 & 12.7 &  4.8 & 0.45 & 109$\times$136 & 40 \\
Abell\,2485          & 22:48:31.13 & $-$16:06:25.6 & 0.247 &  5.9 & 21.2 & 0.19 & 106$\times$126 & 18 \\
Abell\,2631          & 23:37:38.80 & $+$00:16:06.5 & 0.278 &  7.9 & 16.1 & 0.28 & 140$\times$152 & 17 
\enddata
\tablenotetext{a}{J2000. Pointing center for \sza\ observations.}
\tablenotetext{b}{Unflagged, on-source time.}
\tablenotetext{c}{Short, SZ-sensitive baselines only.}
\tablenotetext{d}{Synthesized beam size.}
\end{deluxetable*}

\section{Analysis}
\label{sec:analysis}

\subsection{Weak Gravitational Lensing}
\label{sec:wlAnalysis}
The procedures for determining galaxy cluster masses and related radii
are described in full by \oke; we summarize salient details here. We
adopt \citetalias{OkabeE10}'s ``red+blue'' galaxy sample -- i.e., we
use only those galaxies that are redder or bluer than the cluster red
sequence.  Exclusion of unlensed galaxies in this manner is essential
because cluster/foreground galaxies can bias \Mtfh\ and \Mfh\ low by
$\sim50-20\%$. The mean redshift of the color-selected background
galaxies is estimated by matching their colors and magnitudes to the
COSMOS photometric redshift catalog \citep{IlbertE09}.  The tangential
shear signal is then measured in logarithmically spaced bins, the
measurement in each bin being the error-weighted mean tangential shear
of the galaxies in that bin.  The bins are centered on the brightest
cluster galaxy (BCG) of each cluster.

The spherical mass (or spherical overdensity mass) $M_\Delta$ is
defined as the mass enclosed within a radius $r_\Delta$ for which the
average interior density is $\Delta$ times the critical density of the
universe ($\rho_c\equiv3H(z)^2/8\pi G$).  $M_\Delta$ is estimated by
fitting the measured shear profile to the NFW model prediction
parameterized by $M_\Delta$ and concentration $c_\Delta$, where the
NFW mass profile \citep{NFW} is given as $\rho \propto
r^{-1}(1+c_\Delta r/r_\Delta)^{-2}$.  Both $M_\Delta$ and $c_\Delta$
are allowed to be free in the fit---no mass-concentration relation is
assumed, as is often done in other works
\citep[e.g.,][]{HoekstraE07}---and the $M_\Delta$ values for each
cluster are determined by marginalizing over the posterior
distribution of concentration values. Spherical weak lensing
masses at two of the three overdensities used in this work, $\Delta=$~2500 and 500, are listed in Table~8 of \citetalias{OkabeE10}.
Values for $\Delta=1000$ are calculated from the mass profiles fitted in that work.
Overdensity radii, which are calculable from the masses ($M_\Delta =
4\pi r_\Delta^3\Delta \rho_c /3$), are given in Table~\ref{t:data}.

\subsection{Integrated Sunyaev-Zel'dovich Effect Signal}
\label{sec:szmodel}

For scaling relation analyses, the SZ signal is typically quantified
using the Compton $y$-parameter integrated within a region centered on
a cluster:
\begin{eqnarray}
Y &=& \int y d\Omega = \frac{1}{D_\mathrm{A}^2} \frac{k_B
\sigma_T}{m_\mathrm{e} c^2} \int dl \int n_\mathrm{e} T_\mathrm{e} dA 
\nonumber \\
 &=& \frac{1}{D_\mathrm{A}^2} \frac{\sigma_T}{m_\mathrm{e} c^2} 
\int dl \int P(r) dA .
\label{eq:Y}
\end{eqnarray}
The quantity $YD_\mathrm{A}^2$ is called the intrinsic $y$-parameter,
as it removes the distance dependence of the right side of
Equation~\ref{eq:Y}.

It is customary to integrate Compton-$y$ over the solid angle
subtended by the cluster, as the average change in the CMB temperature
within an aperture ($\overline{\Delta T}\propto Y T_\mathrm{CMB}$) is
expected to be a robust observable.  One standard method of
accomplishing this is to compare the observed sky signal with a
parameterized radial SZ profile, projected to two dimensions and
filtered in the same way as the sky data.  However, both in
interferometric data such as ours and in data from single-aperture
telescopes, the large-scale SZ signal is attenuated by spatial
filtering and/or confused with the anisotropy of the primary CMB
\citep[e.g.,][]{PlaggeE10}.  The poor constraint on the large-scale
behavior of the profile is coupled into the derived $Y$ as additional
uncertainty from the unknown large $l$ (line of sight distance)
behavior of the profile.

This effect can be partially mitigated by adopting as an observable
the integral of the model profile over a spherical volume,
\bq
Y_\mathrm{sph} = \frac{1}{D_\mathrm{A}^2} 
\frac{\sigma_T}{m_\mathrm{e} c^2} \int P(r) dV .
\label{eq:Ys}
\eq
\Ys\ is relatively insensitive to the unconstrained modes in the data,
and can be calculated without resorting to arbitrary outer radius
cutoffs for the line-of-sight integration \citep[as noted
by][]{ArnaudE10}. Furthermore, such integration is standard in the
analysis of X-ray measurements of clusters
\citep[e.g.,][]{ZhangE08,VikhlininE09-CCCP2}.  We therefore perform
our scaling relation analysis using the spherically integrated
intrinsic $y$-parameter, $Y_\mathrm{sph}D_\mathrm{A}^2$.

Since interferometers do not measure the total sky signal, but instead
sample the Fourier transform of the sky signal only at the spatial
frequencies determined by their $uv$ coverage, we must restore missing
spatial information in order to determine our observable.  We
accomplish this by adopting a model profile to fit to our visibility
data; specifically, the generalized NFW (GNFW) pressure profile
proposed by \citet{NagaiE07},
\bq
P(r) = \frac{P_0}{
  x^{\gamma} 
  \left( 1 + x^{\alpha} \right)^{(\beta-\gamma)/\alpha}} ,
\label{eq:P}
\eq
where $x\equiv r/r_\mathrm{s}$. The model contains five parameters:
three exponents ($\alpha$, $\beta$, $\gamma$) which determine the
shape of the profile, a normalization $P_0$, and a scale radius
$r_\mathrm{s}$ (alternatively, a concentration parameter).

Our data are not capable of placing useful constraints on the three
shape parameters due to degeneracies between them, the moderate signal
to noise, and the limited range of angular scales probed by the
interferometer.  We fix the values of the shape parameters to the
\citet{ArnaudE10} average values, [$\alpha$, $\beta$, $\gamma$]=[1.05,
5.49, 0.31], which were determined from simulations and scaled X-ray
pressure profiles of galaxy clusters in the low-redshift ($z<0.2$)
REXCESS sample.
We consider other observationally motivated values of $\alpha$, $\beta$, and $\gamma$
(derived from morphological subsamples in \citealt{ArnaudE10}, see also 
section~\ref{sec:syssz}) to determine the
range of systematic error that may be introduced by the fixed profile shape.
The systematic uncertainty in $Y$ is found to range from 1 to 9\% between \rtfh\ and \rfh\ in our data.

We estimate the remaining pressure model parameters
($P_0$, $r_s$, and the centroid position) and their uncertainties
using a Markov chain Monte Carlo method (MCMC;
\citealt{BonamenteE06}).  At each step of the MCMC, we integrate the
spherical profile along the line of sight, filter it with the primary
beam pattern of the \sza\ antennas (approximately a gaussian with FWHM
11.0\arcmin), and Fourier transform for comparison with the measured
Fourier components (visibilities).  The positions and fluxes of
detected radio sources are included in the MCMC as additional free
parameters, and are marginalized over to determine the pressure model
parameters. 

The centroid of the SZ signal is a free parameter in our model to
allow for the possibility of real offsets between the center of the
ICM and the BCG.  The typical uncertainty in the best-fit SZ centroid
is 0.03\rfh, corresponding to $\sim$10\arcsec.  For all 18 clusters,
the best-fit SZ centroid is within 0.15\rfh\ of the BCG position (the
assumed center of shear), with the SZ centroid in $75\%$ of our sample
lying within $0.04\rfh$ of the respective BCGs.  Larger offsets were
typically found in clusters that lack a cool core, or show signs of
merger activity.  For example, the most significant offset is seen in
A\,2390 ($r_\mathrm{SZ-WL}=0.15\pm0.04$\rfh), which shows both a cool
core and merger activity \citep{AllenE01}.  After accounting for the measurement
uncertainty, the distribution of offsets is reasonably consistent with
recent X-ray/optical studies \citep{SandersonE09,HaarsmaE10}.

To calculate \Ys\ for each cluster, we integrate the respective ICM
model (Equations~\ref{eq:Ys}~\&~\ref{eq:P}) over the spherical volume
defined by the overdensity radii obtained from the weak lensing models
(Section~\ref{sec:wlAnalysis}; Table~\ref{t:data}). These radii are
not constrained from the SZ data alone, so a completely independent
determination of \Ys\ at the overdensities of interest is not
possible.  The correlation between \Mwl\ and \Ys\ introduced by this
procedure is discussed in Section~\ref{sec:scatter}.  Note that the
shared value of $r_\Delta$ is the only coupling between the
measurements of \Ys\ and the lensing analysis---there is no joint
fitting of the mass and pressure profiles.  The final \Ys\
measurements are listed in Table~\ref{t:data}. 
Because of parameter correlations, chiefly between $P_0$ and
$r_\mathrm{s}$, we do not report best-fit values of the individual profile parameters
for each cluster. However, as shown in Section 4.4 of \citet{MroczkowskiE09}, the value of \Ys\
determined from the $P_0$$-$$r_\mathrm{s}$ pairs in the Markov chain is better 
constrained than either of these two parameters alone.

\subsection{Morphological Classification}
\label{sec:morph}

One of our aims is to check whether the shape, normalization, and
intrinsic scatter of the mass-\Ys\ relation depends on cluster
properties.  Of particular interest is whether clusters might exhibit
different behavior based on their dynamical state, which we often
classify simply as merging or non-merging; however, dynamical state is
difficult to estimate from SZ and weak lensing measurements alone.
Here we employ a simple and broadly-accepted indicator of dynamical
state: the centroid shift parameter \ww, defined as the standard
deviation of the projected separation between the X-ray peak and the
centroid of emission calculated in circular apertures centered on the
X-ray peak.

We measure \ww\ for each cluster in our sample using archival
\emph{Chandra} data, following the method of \citet{MohrE95} as
implemented by \citet{MaughanE08}.  The analysis is performed on
exposure-corrected images with sources masked.  The core
($<30{\rm~kpc}$) is excised from the calculation of the centroid, and
the radii span 0.05\rfh\ to \rfh\ in steps of 5\% of \rfh. The core is
included in the calculation of the X-ray peak. The error on \ww\ is
derived from the standard deviation observed in cluster images
resimulated with Poisson noise, following \citet{BohringerE10}.
Previous observational studies of the \ww\ distribution
\citep[e.g.,][]{MaughanE08,PrattE09,BohringerE10} have often divided
clusters into ``disturbed'' and ``undisturbed'' sub-samples at
$\ww=10^{-2}\rfh$. We adopt this value to aid comparison of our
results with the literature.  Six of the 18 clusters are classified as
disturbed ($\ww>10^{-2}\rfh$), and the remaining 12 as undisturbed
($\ww<10^{-2}\rfh$; Table~\ref{t:data})\footnote{Two of the three
  clusters excluded in Section~\ref{sec:sz} have X-ray data as well.
  Of these, one would qualify as disturbed (Abell~115N,
  \ww=7.1$\times10^{-2}$) and the other as undisturbed
  (RX~J1720.1+2638, \ww=0.25$\times10^{-2}$), so their exclusion does
  not preferentially affect one subsample.}. Note that although much of
our discussion makes use of this binary morphological classification
system, equating the disturbed systems with mergers and undisturbed
systems with non-mergers, the true relationship between centroid shift
and dynamical state is likely to be more complex.

\section{Results}
\label{sec:results}

\begin{figure*}
\epsscale{1.15}
\plotone{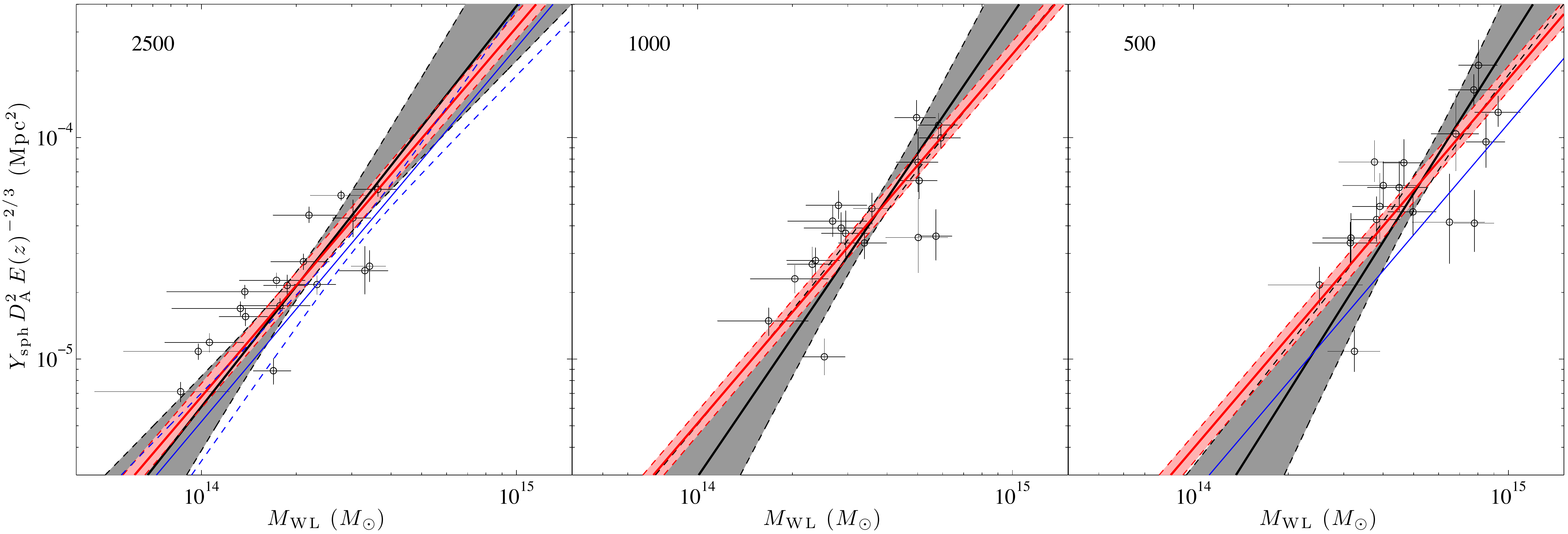}
\caption{The mass-$Y$ scaling relation at three overdensities, shown
  in the upper left corner of each plot. Two fits are shown at each
  overdensity: the best-fit line with a free slope is shown in black,
  and the best-fit line with the slope fixed to self-similar
  (1/\aYM=5/3) is shown in red. The 1$\sigma$ uncertainty regions are
  shaded and marked with dashed lines. Results from prior analyses are
  plotted for comparison with thin blue lines: the $M$$-$\Ys\ scaling relation
  derived from the data of \citet{BonamenteE08} (left panel), and the
  scaling relation from \citet{AnderssonE11} (right panel).}
\label{f:MY}
\end{figure*}

The weak-lensing mass and \Ys\ measurements are compared in
Figure~\ref{f:MY}, and a positive correlation is evident at all radii.
In this section, we define the scaling relation model that we will fit
to the data, describe the regression analysis, and present the main
results.

\begin{deluxetable*}{lcccccccccc}
\tablecolumns{11}
\tablewidth{0pt}
\tablecaption{Cluster Mass and \Ys\label{t:data}}
\tablehead{\scriptsize Cluster & \scriptsize\ww\ &
\scriptsize$r_{2500}$ & \scriptsize$M_{2500}$ & \scriptsize$Y_\mathrm{sph,2500}D_\mathrm{A}^2$ &
\scriptsize$r_{1000}$  & \scriptsize$M_{1000}$ & \scriptsize$Y_\mathrm{sph,1000}D_\mathrm{A}^2$ & 
\scriptsize$r_{500}$  & \scriptsize$M_{500}$ & \scriptsize\hfil$Y_\mathrm{sph,500}D_\mathrm{A}^2$\hfil \\
 & \scriptsize($10^{-2}$~$r_{500}$) &
\scriptsize(Mpc) & \scriptsize($10^{14}$~\Msun) & \scriptsize($10^{-5}$ Mpc$^2$) & 
\scriptsize(Mpc) & \scriptsize($10^{14}$~\Msun) & \scriptsize($10^{-5}$ Mpc$^2$) & 
\scriptsize(Mpc) & \scriptsize($10^{14}$~\Msun) & \scriptsize($10^{-5}$ Mpc$^2$)}
\startdata
\scriptsize Abell 68         &\scriptsize 1.17$\pm$0.14 &\scriptsize 0.41 &\scriptsize  1.37$^{+0.57}_{-0.60}$ &\scriptsize  2.18$^{+0.16}_{-0.15}$ &\scriptsize 0.70 &\scriptsize  2.69$^{+0.76}_{-0.75}$ &\scriptsize  4.54$^{+0.81}_{-0.67}$ &\scriptsize 1.01 &\scriptsize  4.01$^{+1.18}_{-1.03}$ &\scriptsize  6.57$^{+1.86}_{-1.39}$ \\
\scriptsize Abell 209        &\scriptsize 0.68$\pm$0.10 &\scriptsize 0.48 &\scriptsize  2.11$^{+0.44}_{-0.45}$ &\scriptsize  2.93$^{+0.25}_{-0.23}$ &\scriptsize 0.88 &\scriptsize  5.05$^{+0.71}_{-0.69}$ &\scriptsize  6.80$^{+1.47}_{-1.17}$ &\scriptsize 1.32 &\scriptsize  8.49$^{+1.26}_{-1.15}$ &\scriptsize 10.18$^{+3.44}_{-2.39}$ \\
\scriptsize RXC J0142.0+2131 &\scriptsize 0.76$\pm$0.13 &\scriptsize 0.45 &\scriptsize  1.87$^{+0.30}_{-0.30}$ &\scriptsize  2.35$^{+0.27}_{-0.27}$ &\scriptsize 0.72 &\scriptsize  2.95$^{+0.52}_{-0.48}$ &\scriptsize  4.03$^{+1.07}_{-0.88}$ &\scriptsize 0.99 &\scriptsize  3.91$^{+0.80}_{-0.71}$ &\scriptsize  5.33$^{+2.18}_{-1.53}$ \\
\scriptsize Abell 267        &\scriptsize 1.68$\pm$0.05 &\scriptsize 0.42 &\scriptsize  1.38$^{+0.24}_{-0.24}$ &\scriptsize  1.67$^{+0.15}_{-0.15}$ &\scriptsize 0.67 &\scriptsize  2.31$^{+0.41}_{-0.39}$ &\scriptsize  2.88$^{+0.56}_{-0.50}$ &\scriptsize 0.94 &\scriptsize  3.16$^{+0.66}_{-0.59}$ &\scriptsize  3.78$^{+1.10}_{-0.88}$ \\
\scriptsize Abell 383        &\scriptsize 0.20$\pm$0.04 &\scriptsize 0.45 &\scriptsize  1.69$^{+0.23}_{-0.23}$ &\scriptsize  0.94$^{+0.13}_{-0.12}$ &\scriptsize 0.70 &\scriptsize  2.53$^{+0.41}_{-0.38}$ &\scriptsize  1.08$^{+0.22}_{-0.19}$ &\scriptsize 0.96 &\scriptsize  3.24$^{+0.66}_{-0.57}$ &\scriptsize  1.15$^{+0.29}_{-0.22}$ \\
\scriptsize Abell 521        &\scriptsize 5.16$\pm$0.12 &\scriptsize 0.38 &\scriptsize  1.06$^{+0.30}_{-0.29}$ &\scriptsize  1.28$^{+0.12}_{-0.12}$ &\scriptsize 0.67 &\scriptsize  2.37$^{+0.46}_{-0.45}$ &\scriptsize  3.01$^{+0.43}_{-0.39}$ &\scriptsize 0.99 &\scriptsize  3.81$^{+0.70}_{-0.65}$ &\scriptsize  4.60$^{+1.23}_{-0.94}$ \\
\scriptsize Abell 586        &\scriptsize 0.26$\pm$0.06 &\scriptsize 0.57 &\scriptsize  3.30$^{+0.61}_{-0.57}$ &\scriptsize  2.64$^{+0.76}_{-0.57}$ &\scriptsize 0.89 &\scriptsize  5.02$^{+1.21}_{-1.06}$ &\scriptsize  3.72$^{+1.85}_{-1.14}$ &\scriptsize 1.22 &\scriptsize  6.50$^{+1.89}_{-1.54}$ &\scriptsize  4.37$^{+2.85}_{-1.52}$ \\
\scriptsize Abell 611        &\scriptsize 0.33$\pm$0.08 &\scriptsize 0.44 &\scriptsize  1.78$^{+0.44}_{-0.45}$ &\scriptsize  1.91$^{+0.15}_{-0.15}$ &\scriptsize 0.75 &\scriptsize  3.39$^{+0.60}_{-0.58}$ &\scriptsize  3.66$^{+0.63}_{-0.56}$ &\scriptsize 1.07 &\scriptsize  4.98$^{+0.91}_{-0.84}$ &\scriptsize  5.06$^{+1.32}_{-1.08}$ \\
\scriptsize Abell 697        &\scriptsize 0.58$\pm$0.09 &\scriptsize 0.48 &\scriptsize  2.19$^{+0.50}_{-0.51}$ &\scriptsize  4.88$^{+0.44}_{-0.37}$ &\scriptsize 0.85 &\scriptsize  4.96$^{+0.74}_{-0.73}$ &\scriptsize 13.43$^{+2.66}_{-2.10}$ &\scriptsize 1.26 &\scriptsize  8.04$^{+1.18}_{-1.09}$ &\scriptsize 23.16$^{+7.00}_{-5.06}$ \\
\scriptsize Abell 1835       &\scriptsize 0.22$\pm$0.01 &\scriptsize 0.52 &\scriptsize  2.77$^{+0.55}_{-0.56}$ &\scriptsize  5.92$^{+0.30}_{-0.29}$ &\scriptsize 0.91 &\scriptsize  5.92$^{+0.92}_{-0.87}$ &\scriptsize 10.76$^{+1.35}_{-1.11}$ &\scriptsize 1.33 &\scriptsize  9.28$^{+1.64}_{-1.46}$ &\scriptsize 14.05$^{+2.59}_{-1.95}$ \\
\scriptsize ZwCl 1454.8+2233 &\scriptsize 0.37$\pm$0.02 &\scriptsize 0.35 &\scriptsize  0.86$^{+0.37}_{-0.40}$ &\scriptsize  0.77$^{+0.08}_{-0.08}$ &\scriptsize 0.60 &\scriptsize  1.68$^{+0.56}_{-0.52}$ &\scriptsize  1.61$^{+0.24}_{-0.23}$ &\scriptsize 0.86 &\scriptsize  2.51$^{+0.93}_{-0.78}$ &\scriptsize  2.34$^{+0.48}_{-0.44}$ \\
\scriptsize ZwCl 1459.4+4240 &\scriptsize 1.29$\pm$0.16 &\scriptsize 0.44 &\scriptsize  1.73$^{+0.40}_{-0.41}$ &\scriptsize  2.48$^{+0.20}_{-0.19}$ &\scriptsize 0.70 &\scriptsize  2.80$^{+0.64}_{-0.59}$ &\scriptsize  5.42$^{+0.89}_{-0.71}$ &\scriptsize 0.97 &\scriptsize  3.76$^{+0.97}_{-0.86}$ &\scriptsize  8.50$^{+2.11}_{-1.56}$ \\
\scriptsize Abell 2219       &\scriptsize 1.75$\pm$0.03 &\scriptsize 0.58 &\scriptsize  3.63$^{+0.57}_{-0.60}$ &\scriptsize  6.25$^{+0.47}_{-0.45}$ &\scriptsize 0.91 &\scriptsize  5.83$^{+0.87}_{-0.81}$ &\scriptsize 12.20$^{+1.58}_{-1.42}$ &\scriptsize 1.27 &\scriptsize  7.77$^{+1.44}_{-1.31}$ &\scriptsize 17.58$^{+3.07}_{-2.61}$ \\
\scriptsize Abell 2261       &\scriptsize 0.81$\pm$0.07 &\scriptsize 0.56 &\scriptsize  3.42$^{+0.42}_{-0.43}$ &\scriptsize  2.81$^{+0.50}_{-0.42}$ &\scriptsize 0.91 &\scriptsize  5.72$^{+0.71}_{-0.68}$ &\scriptsize  3.84$^{+1.23}_{-0.85}$ &\scriptsize 1.27 &\scriptsize  7.80$^{+1.18}_{-1.08}$ &\scriptsize  4.40$^{+1.80}_{-1.13}$ \\
\scriptsize RXC J2129.6+0005 &\scriptsize 0.43$\pm$0.11 &\scriptsize 0.41 &\scriptsize  1.33$^{+0.51}_{-0.52}$ &\scriptsize  1.82$^{+0.13}_{-0.14}$ &\scriptsize 0.72 &\scriptsize  2.86$^{+0.66}_{-0.68}$ &\scriptsize  4.19$^{+0.73}_{-0.60}$ &\scriptsize 1.05 &\scriptsize  4.50$^{+1.05}_{-0.94}$ &\scriptsize  6.39$^{+2.01}_{-1.37}$ \\
\scriptsize Abell 2390       &\scriptsize 0.78$\pm$0.03 &\scriptsize 0.54 &\scriptsize  3.03$^{+0.43}_{-0.42}$ &\scriptsize  4.64$^{+0.97}_{-0.83}$ &\scriptsize 0.87 &\scriptsize  5.01$^{+0.79}_{-0.74}$ &\scriptsize  8.29$^{+3.10}_{-2.19}$ &\scriptsize 1.21 &\scriptsize  6.81$^{+1.24}_{-1.13}$ &\scriptsize 11.15$^{+5.75}_{-3.55}$ \\
\scriptsize Abell 2485       &\scriptsize 0.53$\pm$0.14 &\scriptsize 0.37 &\scriptsize  0.98$^{+0.40}_{-0.41}$ &\scriptsize  1.17$^{+0.10}_{-0.10}$ &\scriptsize 0.64 &\scriptsize  2.04$^{+0.57}_{-0.57}$ &\scriptsize  2.48$^{+0.37}_{-0.34}$ &\scriptsize 0.93 &\scriptsize  3.15$^{+0.86}_{-0.76}$ &\scriptsize  3.61$^{+0.87}_{-0.71}$ \\
\scriptsize Abell 2631       &\scriptsize 1.68$\pm$0.14 &\scriptsize 0.49 &\scriptsize  2.33$^{+0.34}_{-0.36}$ &\scriptsize  2.36$^{+0.23}_{-0.23}$ &\scriptsize 0.76 &\scriptsize  3.58$^{+0.48}_{-0.45}$ &\scriptsize  5.21$^{+0.91}_{-0.78}$ &\scriptsize 1.05 &\scriptsize  4.66$^{+0.70}_{-0.67}$ &\scriptsize  8.37$^{+2.30}_{-1.80}$ 

\enddata
\tablecomments{Cluster masses at $\Delta=2500$ and 500 as reported in \citetalias{OkabeE10} for NFW halo fits to
the shear profile, $\Delta=1000$ derived from the profile fits used in that work. An additional systematic uncertainty of 
10\% in the overall scaling of the $Y$ values is not included in the errors reported here.}
\end{deluxetable*}

\subsection{Self-similar Scaling}
Without significant influence from non-gravitational processes, the
mass, temperature, size, and other properties of galaxy clusters are
expected to follow self-similar scaling relations \citep{Kaiser86}.
These power-law relationships follow directly from the assumptions of
hydrostatic equilibrium and an isothermal distribution of baryons and
dark matter \citep[e.g.,][]{BryanNorman98}, and form a useful
reference point for measurements of scaling relations.

The simplest self-similar model, as derived in the above-cited papers,
predicts that cluster mass ($M$) and temperature ($T$) are related as
\begin{equation}
  T\propto M^{2/3} E(z)^{2/3} \Delta^{1/3} .
\label{eq:TM}
\end{equation}
To derive the expected scaling between mass and the SZ
signal, we note that the double integral on the right hand side of
Equation~\ref{eq:Y} can be written as $M_\mathrm{gas}T_\mathrm{e}$ for
an isothermal ICM, or equivalently as $f_{gas} M T_\mathrm{e}$ for
$f_{gas}\equiv M_{gas}/M$. Combining this with Equation~\ref{eq:TM},
we find the \my\ scaling relation:
\begin{equation}
  M \propto f_{gas}^{-3/5}  \Delta^{-1/5}  \left[YD_\mathrm{A}^2 E(z)^{-2/3}\right]^{3/5} . 
\label{eq:MY}
\end{equation}
This equation defines the self-similar reference model for our scaling
analysis.

\subsection{Regression}
We analyze the scaling between mass and \Ys\ at three overdensity
radii ($\Delta$~=~500, 1000, and 2500) determined from the weak lensing
measurements. We fit the data with a power law of the form
\begin{equation}
  \frac{M(r_\Delta)}{10^{14} M_\odot}=
  10^A\left(\frac{Y_\mathrm{sph}D_\mathrm{A}^2E(z)^{-2/3}}{10^{-5}\mathrm{Mpc}^2}\right)^B .
\label{eq:scale}
\end{equation}
The regression is performed in linearized coordinates by using the
base-10 logarithm of the data points.

Our data have significant variations in their measurement precision,
and simulations suggest that the \my\ correlation should have
intrinsic scatter, placing significant demands on our fitting
technique.  The problem of linear regression with uncertainty in both
axes and intrinsic scatter is complex, and several methods of
parameter estimation have been formulated for this circumstance
\citep[e.g.,][]{AkritasBershady96,TremaineE02,WeinerE06}. Methods that
ignore the intrinsic scatter, particularly when accompanied by
heterogeneous measurement errors, lead to biased regression parameters
\citep{Kelly07}.

We perform our linear regression using a three-parameter model:
normalization $A$, slope $B$, and intrinsic scatter \SigYM.  Parameter
determinations are achieved through the Bayesian regression method
presented in \citet{Kelly07}.  In that work \citeauthor{Kelly07}
demonstrates that this method performs well in regimes where parameter
estimates from other methods (OLS, BCES, FITEXY) are significantly
biased.  The regression is performed with publicly available IDL code
written by \citeauthor{Kelly07}.

\subsection{Scaling Relations}
\label{s:scaleR}
\begin{deluxetable}{lccccc}
\tablecolumns{6}
\tablewidth{0pt}
\tabletypesize{\scriptsize}
\tablecaption{\Mwl$-$\Ys\ Scaling Relations\label{t:MY}}
\tablehead{ & Cluster &  & 
& & \SigYM\tablenotemark{c} \\
$\Delta$  & Subset\tablenotemark{a} & $A$ & $B$ & Cov($AB$)\tablenotemark{b} & (\%) }
\startdata
 500 & All &  0.367$^{+0.096}_{-0.099}$ &  0.44$^{+ 0.12}_{- 0.11}$ & -0.012 &   21$^{+  9}_{-  8}$ \\ 
1000 & All &  0.254$^{+0.077}_{-0.081}$ &  0.48$^{+ 0.11}_{- 0.11}$ & -0.009 &   19$^{+  9}_{-  7}$ \\ 
2500 & All &  0.118$^{+0.060}_{-0.066}$ &  0.55$^{+ 0.14}_{- 0.13}$ & -0.008 &   20$^{+  9}_{-  7}$ \\ 
\hline 
 500 & All &  0.241$^{+0.036}_{-0.036}$ & 3/5 & \nodata &   20$^{+ 10}_{-  8}$ \\ 
 500 & U   &  0.297$^{+0.043}_{-0.050}$ & 3/5 & \nodata &   24$^{+ 13}_{- 10}$ \\ 
 500 & D   &  0.146$^{+0.062}_{-0.071}$ & 3/5 & \nodata &   17$^{+ 21}_{-  9}$ \\ 
1000 & All &  0.174$^{+0.031}_{-0.032}$ & 3/5 & \nodata &   19$^{+  9}_{-  7}$ \\ 
1000 & U   &  0.212$^{+0.040}_{-0.043}$ & 3/5 & \nodata &   24$^{+ 12}_{-  9}$ \\ 
1000 & D   &  0.104$^{+0.053}_{-0.055}$ & 3/5 & \nodata &   15$^{+ 18}_{-  8}$ \\ 
2500 & All &  0.101$^{+0.031}_{-0.035}$ & 3/5 & \nodata &   19$^{+  9}_{-  7}$ \\ 
2500 & U   &  0.116$^{+0.043}_{-0.048}$ & 3/5 & \nodata &   25$^{+ 13}_{-  9}$ \\ 
2500 & D   &  0.062$^{+0.062}_{-0.069}$ & 3/5 & \nodata &   22$^{+ 24}_{- 11}$ 

\enddata
\tablecomments{Scaling relations fitted according to
Equation~\ref{eq:scale}. For fits in the lower section of the table the power law slope was fixed to the self-similar value (\aYM=3/5).}
\tablenotetext{a}{Cluster sample used in fitting. All clusters,
(U) undisturbed, or (D) disturbed.}
\tablenotetext{b}{The covariance of the parameters $A$ and $B$. Because the scaling relation pivot point is not optimized, this quantity is non-zero and should be incorporated in any evaluation of the scaling relation uncertainty.}
\tablenotetext{c}{Though the regression is performed in base-10 logarithmic space, these numbers are converted to base-$e$ and are therefore percentage scatter.}
\end{deluxetable}

We fit the scaling relation (Equation~\ref{eq:scale}) to the full
sample of clusters with the slope $B$ as a free parameter, referring
to the result as the ``free slope'' (FS) fit (Table~\ref{t:MY}).  We
also report results with the slope fixed to the self-similar value of
$B=0.6$, which we refer to as the ``self-similar'' (SS) fit.
The posterior distributions of the normalization and scatter in the SS
fit are obtained by taking links with the appropriate slope from the
Markov chains generated by the fitting routine.

At all three overdensity radii, the FS mass-\Ys\ relations are
slightly shallower than the self-similar prediction of $B=0.6$
(Figure~\ref{f:MY}), although the differences from self-similarity are
not statistically significant; the relation at $\Delta=500$ is the
most discrepant at $1.4\sigma$ significance. A jackknife test of the
\rfh\ data, dropping one cluster at a time and refitting the
17-cluster scaling relations, shows that the cluster with the lowest
$Y_{\mathrm{sph,}500}$ (Abell~383) affects the slope far more than any
other. Without this cluster the \rfh\ scaling relation would have
parameters $A=0.259^{+0.131}_{-0.138}$, $B=0.56^{+0.16}_{-0.15}$,
\SigYM$=20^{+10}_{-8}$\%; in this case $B$ is in no tension with the
self-similar prediction.  We have no strong reason to exclude this
cluster, and the removal of outliers may bias the average scaling
relation parameters, so no outlier clusters are excluded from the scaling relations
presented in this paper.

The scaling relations reported in Table~\ref{t:MY} are subject to some bias because of  
the $L_\mathrm{X}$ selection threshold of the parent LoCuSS sample. If 
there is a correlation between the scatter in the two ICM observables, 
$L_\mathrm{X}$ and \Ys\ (or \Mwl), selection on the former will affect the 
distribution of the latter. Such correlation is not yet well-characterized 
observationally, however. The possibility of systematic effects in the 
scatter (Section~\ref{sec:scatter}), the previously noted influence of 
outliers in this small sample, and the possible bias from the X-ray selection
demand caution when employing these scaling relations for other analyses. 

The intrinsic scatter,
$\sigma_{M|Y}$, is found to be $\sim 20$\% at all overdensity radii,
which is larger than the $\sim 10$\% predicted in
numerical simulations \citep[e.g.,][]{Nagai06}.  This holds for the SS
fits as well, confirming that the measured scatter is not strongly
dependent on the slope parameter.

The low scatter reported for simulated $Y$$-$$M$ scaling relations
does not capture additional complications introduced by the
measurement techniques.  One potentially important source of scatter
in our data is the contribution to \Ys\ from lower mass haloes along
the line of sight.  This was examined by \citet{HolderE07} and found
to be insignificant for our mass and redshift range.  The scatter
between the weak-lensing-derived mass and true halo mass may also be
important. In measurements of the type used here it is expected to be
$\sim20$\% \citep[e.g.,][]{BeckerKravtsov11}, which is in good
agreement with the scatter we observe.

\section{Discussion}
\label{sec:discussion}

\subsection{Comparison to Previous Results}
\label{sec:prev}

The \my\ scaling relations presented above are the first to combine SZ
measurements with weak gravitational lensing.  Previous observational
studies have assumed HSE to derive mass measurements, and consequently
have different sets of systematics and biases.  The assumption of HSE
introduces an intrinsic correlation between mass and \Ys, while the
intrinsic correlation between the lensing-based mass and \Ys\ in our
study is expected to be lower (although not necessarily zero, as
discussed in \S\ref{sec:correlation}).  Moreover, systematic biases
between HSE-based masses and lensing-based masses have been measured
\citep{MahdaviE08} or limited at the 10\% level
\citep{ZhangE08,ZhangE10} in cluster observations, and found between
HSE-based masses and true masses in simulated clusters
\citep{RasiaE06,LauE09,BurnsE10,CavaliereE11}.  These simulations
suggest that hydrostatic masses are biased low by $\sim10-20\%$ at
\rfh, with smaller biases at higher over-densities.  In order to
explore these effects, we compare our results with previous
observational and theoretical studies in \S\ref{sec:prevobs} and
\S\ref{sec:prevsim}, respectively.

\subsubsection{Previous Observational Results}\label{sec:prevobs}

Our results are consistent within $\sim1\sigma$ with those of
\citet{BonamenteE08} at $\Delta=2500$ (left panel of
Figure~\ref{f:MY}).  \citeauthor{BonamenteE08}\ analysed \ovro\ and
\bima\ observations of 19 clusters at $0.14<z<0.3$ for which archival
\chandra\ data were available.  The clusters were modeled with an
isothermal $\beta$-model (after excising the central 100~kpc of the
X-ray data) and masses were determined assuming HSE. The $Y$ values
were determined by normalizing the shape of the $\beta$-model density
(or pressure, since the cluster is assumed isothermal) profile with
the SZ measurements and integrating. We calculate the \Ys\ values from
the Markov chains used in that paper and perform the linear regression
using the methods described above. Their results imply
$A=0.131\pm0.074$, $\aYM=0.65\pm0.12$, $\SigYM=9^{+7}_{-5}\%$, which
are consistent with self-similarity, and are statistically
indistinguishable from our best fit parameters (Table~\ref{t:MY}).
Despite the lack of statistical significance, it is interesting to
note that the intrinsic scatter in \citeauthor{BonamenteE08}'s
HSE-based scaling relation is roughly half that found in our
lensing-based relation.

Our results at $\Delta=500$ agree less well with \citet{AnderssonE11}
(right panel of Figure~\ref{f:MY}), who examined the $\Mfh-\Ys$
scaling relation for a sample of galaxy clusters discovered by the SPT
\citep{VanderlindeE10}. The mass range of the \citet{AnderssonE11} sample
is similar to that presented here, though at higher redshift (median $z=0.74$). 
They estimated cluster masses from the
pressure-like X-ray observable $Y_X$ through an observational
calibration of the $\Mfh-Y_X$ scaling relation, with \Mfh\ measured
from X-ray data assuming HSE \citep{VikhlininE09-CCCP2}. The SZ
profile, largely unresolved by the SPT, was modeled using the observed
X-ray density profile and a universal temperature profile
\citep{VikhlininE06}.  The resulting scaling relation
($A=0.36\pm0.03$, $\aYM=0.60^{+0.12}_{-0.09}$, $\SigYM=9\pm5\%$, in
our nomenclature) is slightly discrepant with our FS model. However,
comparison with our self-similar model is more direct, as
\citeauthor{AnderssonE11}'s slope matches the self-similar value.  The
scatter is again roughly half that observed in our data. Their
normalization is $\sim30\%$ higher in mass at fixed \Ys\ than our
self-similar fit, a $2.3\sigma$ difference (after accounting for the
contribution of the intrinsic scatter, compare the red and blue lines
on the right panel of Fig~\ref{f:MY}).  This difference is not
attributable to the hydrostatic mass bias discussed above
(\S\ref{sec:prev}), because the assumption of HSE results in a bias in
the direction opposite to the observed discrepancy.  The origin of
this difference remains unclear, but systematic effects in our data, discussed below, 
may play a significant role.

\subsubsection{Previous Simulations}\label{sec:prevsim}

\begin{figure}
\epsscale{1.15}
\plotone{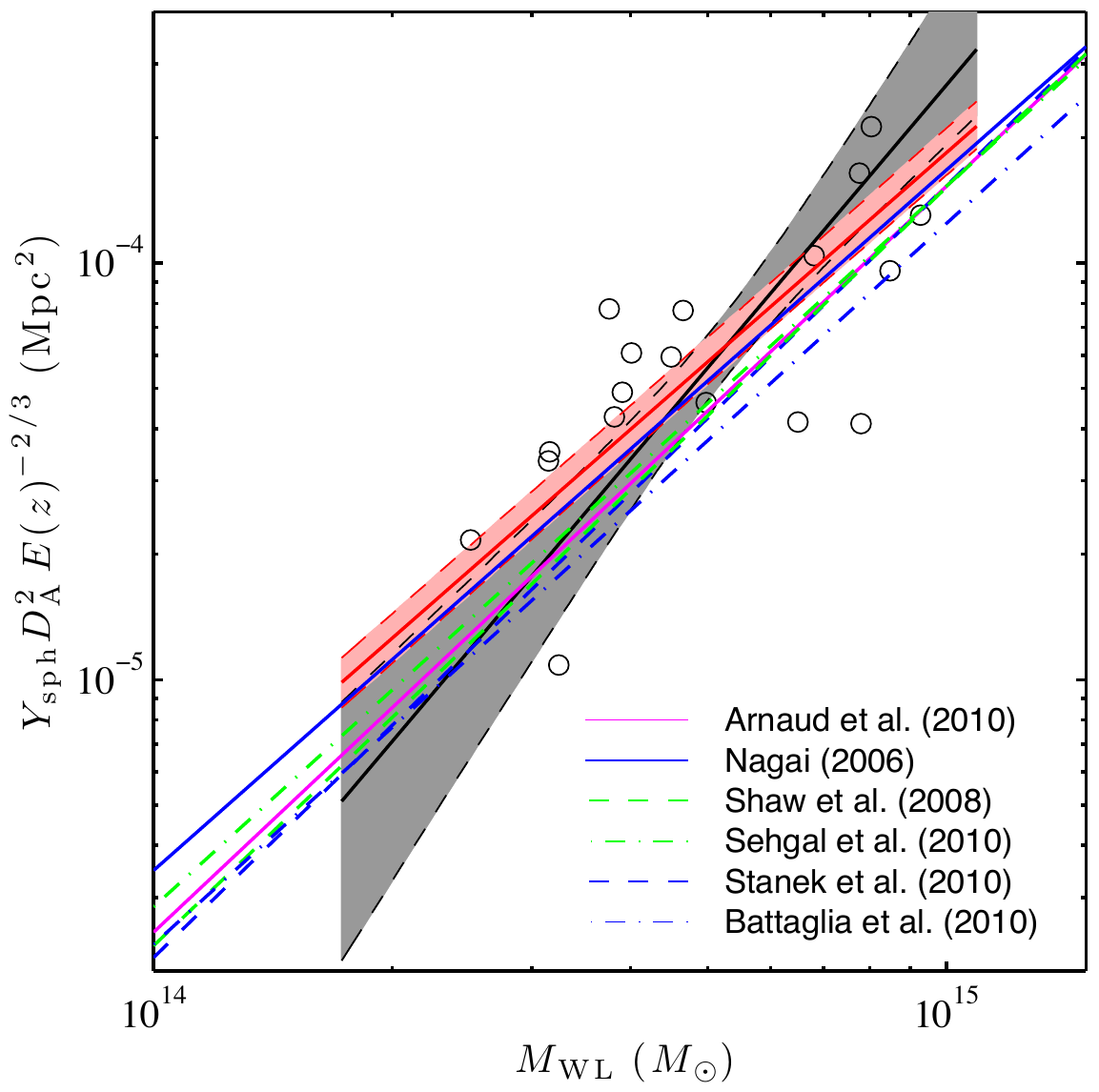}
\caption{Data and scaling relations at $\Delta=500$ compared to
  scaling relations measured from X-ray data and numerical simulations
  of galaxy clusters. The scaling relation fits, plotted as in
  Figure~\ref{f:MY}, are truncated at the boundaries of the mass range
  of the cluster sample including 1$\sigma$ uncertainty. Data points
  from this work are shown as open circles without error bars.  Five
  simulated scaling relations between \Ys\ and true \Mfh\ are shown,
  as well as the scaling relation predicted by \citet{ArnaudE10} from
  X-ray observations and numerical simulations.  The \citet{Nagai06}
  scaling relation was derived from simulated clusters with mass
  greater than $10^{14}$~\Msun, while the \citet{SehgalE10} relation
  was determined from simulated \Mfh$>2\times10^{14}$~\Msun clusters.}
\label{f:simMY}
\end{figure}

Accurate calibration of numerical simulations is a key step towards
cosmological application of SZ surveys.  Many authors have therefore
developed simulations of increasing sophistication, five recent
examples of which are over-plotted on our data in
Figure~\ref{f:simMY}.  In all cases we show the predicted scaling
relation between mass and $\Ys$, using fits provided by the authors
for those simulation studies that did not publish a mass-$\Ys$
relation.  The simulated relations are very close to self-similar
($B=0.6$; Table~\ref{t:YMsim}), and slightly under-predict $\Ys$ at
fixed mass relative to our best-fit self-similar model (red line in
Figure~\ref{f:simMY}).  The fractional offset in $\Ys$ of the
simulated relations from the observed relation is $\sim10-35\%$ at
$\Mfh=5\times10^{14}$\Msun, the mean mass of the observed sample.  We
also rank the simulations using a simple $\chi^2$ comparison of their
predicted scaling relations with the observational data, assuming an
intrinsic scatter of 20\% in mass at fixed $\Ys$
(Table~\ref{t:YMsim}).  We find that the \citet[][blue solid line in
Figure~\ref{f:simMY}]{Nagai06} simulation is the closest match to the
observations, and the \citet{BattagliaE10} simulation is the poorest
match, though not even $\sim2\sigma$ deviant from our self-similar
relation.  Note that this ranking does not depend on the adopted
intrinsic scatter, although the absolute value of $\chi^2$ does.

\begin{deluxetable}{lccll}
\tablecolumns{5}
\tablewidth{0pt}
\tablecaption{Simulated Mass$-$\Ys\ Scaling Relations\label{t:YMsim}}
\tablehead{Ref\tablenotemark{a} & $A$\tablenotemark{b} & $B$\tablenotemark{b} & Notes & $\chi^2~$\tablenotemark{c}}
\startdata
(1) &  0.27 & 0.60 & $M>10^{14}$\Msun, $z=0,0.6$ & 15.8 \\
(2) &  0.35 & 0.55 & & 18.9 \\
(3) &  0.32 & 0.58 & $M>1.5\times10^{14}$\Msun, $0.1\le z \le 0.2$ & 17.8 \\
(4) &  0.36 & 0.54 & Preheating Model &  19.1 \\
(5) &  0.37 & 0.58 & AGN Feedback Model & 26.0 
\enddata
\tablenotetext{a}{(1) \citet{Nagai06}; (2) \citet{ShawE08}; (3)
  \citet{SehgalE10}; (4) \citet{StanekE10}; (5) \citet{BattagliaE10}.}
\tablenotetext{b}{Scaling relations parameterized by $A$ and $B$ as in
  Equation~\ref{eq:scale}.}  
\tablenotetext{c}{$\chi^2$ between model and the 18 data points (17
  d.o.f) for an assumed intrinsic scatter of 20\%.}
\end{deluxetable}

In a hybrid observational and simulated calibration of the \my\
relation, \citet{ArnaudE10} constructed an ``observed'' scaling
relation between \Ys\ and \Mfh\ without SZ measurements.  They use the
universal pressure profile described in Section~\ref{sec:szmodel},
which they calibrate against X-ray pressure profiles and simulated
clusters, inside and outside of \rfh, respectively.  Their best-fit
relation ($A$=0.34, $B$=0.56) also appears in Figure~\ref{f:simMY}. It
matches well with the simulations of \citet{ShawE08} and
\citet{SehgalE10}, similarly underpredicting $Y$ at fixed $M$. The
mass scale for this scaling relation comes from their $M$$-$$Y_X$
scaling relation, which is calibrated against hydrostatic masses, so
the line might instead have been expected to over-predict $Y$ due to
the hydrostatic bias. At the measurement precision and level of
scatter in this work we do not detect such a bias, even though it is
expected to be more important at \rfh\ than at \rtfh\ where we are
able to compare with the results of \citet{BonamenteE08}.

\subsection{Observed Scatter and Morphological Segregation}
\label{sec:scatter}

\begin{figure}
  \epsscale{1.15}
  \plotone{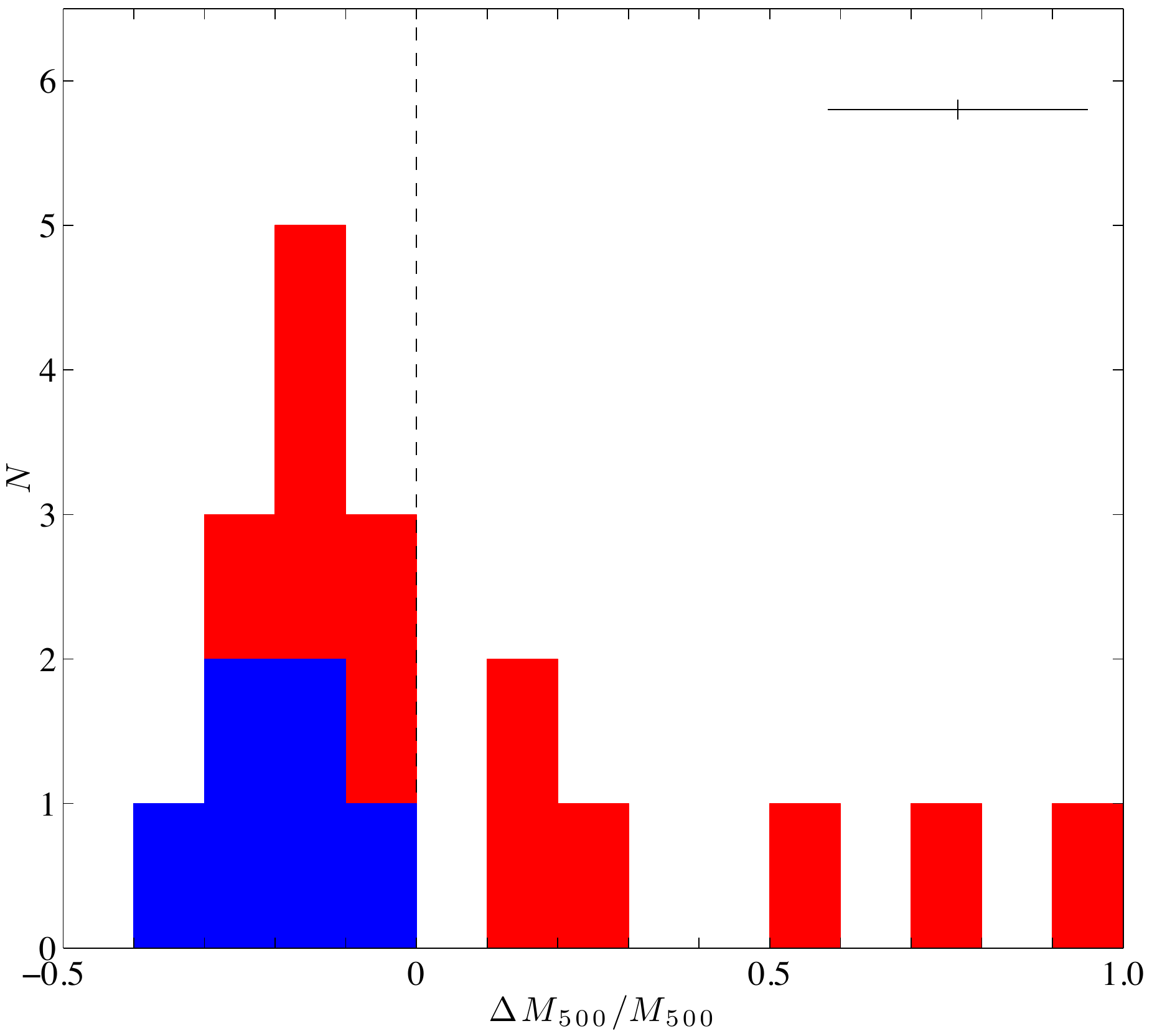}
  \caption{Distribution of fractional deviation in mass at
    $\Delta=500$ of the 18 clusters in our sample from the best-fit
    self-similar scaling relation fit shown as the solid red line in
    the right panel of Fig.~\ref{f:MY}.  Disturbed clusters (blue;
    defined in \S\ref{sec:morph}) lie exclusively on the low-mass side
    of the mean relation; undisturbed clusters (red) lie on both sides
    of the mean relation, with a tail to large positive deviations. A typical
    error bar for the fractional deviation is show in the upper right. }
  \label{f:dev}
\end{figure}

\begin{figure*}
\epsscale{1.15}
\plotone{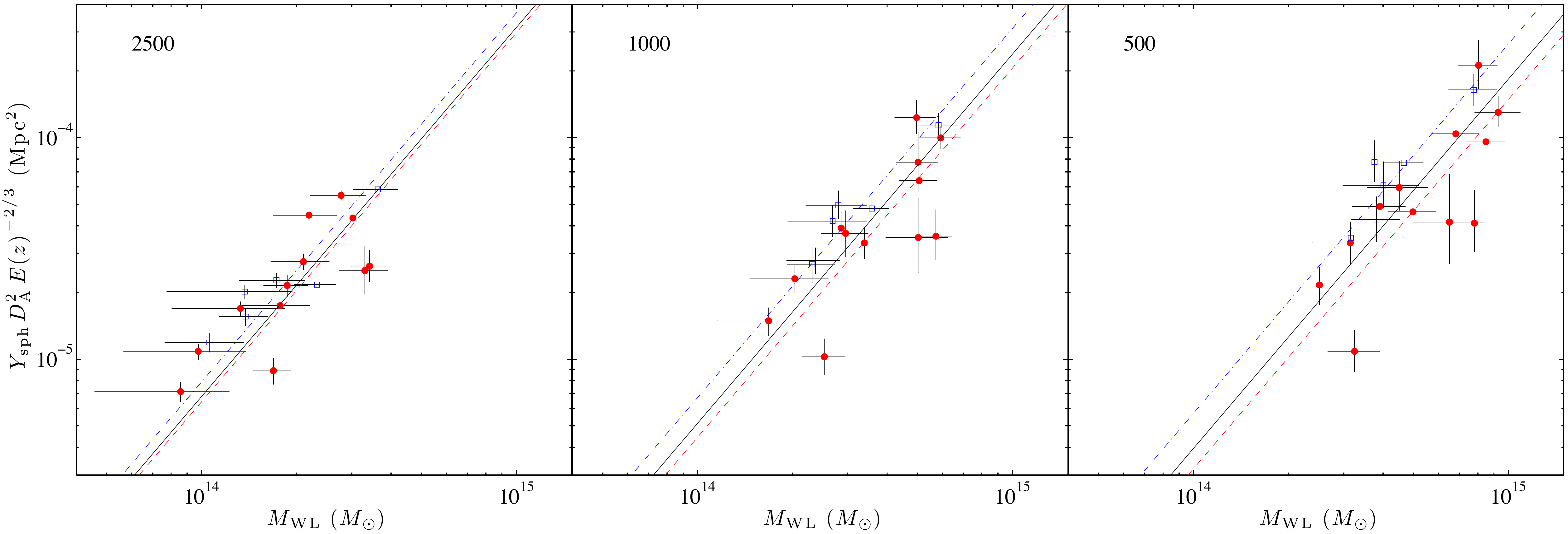}
\caption{The mass-$Y$ scaling relation at each overdensity, with
  cluster morphology indicated by the data symbols. Disturbed clusters
  are marked with open squares and undisturbed clusters with filled
  circles. The self-similar fit for the full sample is shown with a
  solid (black) line, and the self-similar fits to the undisturbed
  (red dashed) and disturbed (blue dot-dashed) subsamples are also
  shown.}
\label{f:morphMY}
\end{figure*}

The distribution of fractional deviations in mass ($\Delta
M_\Delta/M_\Delta$) of clusters from the mean $M_\Delta-\Ys$ relations
is asymmetric, particularly at $\Delta=500$ (Figure~\ref{f:dev}),
where it peaks to the low-mass side of the relations (negative
deviations) and has an extended tail to the high-mass side (positive
deviations).  Interestingly, the clusters classified as ``disturbed''
on the basis of the centroid shift measurements discussed in
\S\ref{sec:morph} all have $\Delta\Mfh/\Mfh<0$, and the tail of the
distribution at $\Delta\Mfh/\Mfh>0.3$ comprises solely undisturbed
clusters (compare the blue and red in Figure~\ref{f:dev}).  This
suggests that the large intrinsic scatter in our mass-\Ys\ relations
may be related to cluster morphology.

We detect a significant difference in the normalization of the 
self-similar mass-\Ys\ relations between disturbed and undisturbed 
clusters at all three overdensity radii.  At fixed \Ys, the mass of 
undisturbed clusters exceeds that of disturbed clusters by $13\pm6\%$, 
$28\pm5\%$, and $41\pm6\%$ at $\Delta=2500$, 1000, and 500, respectively 
(Table~\ref{t:MY}; Figure~\ref{f:morphMY}).  
The probability of observing such an offset randomly is low even in scattered data like these; drawing random samples (with replacement) of 6 and 12 clusters from our data and repeating our fits, we find that only $1\%$ of random sub-samples are more significantly offset than the real data at \rfh.

If confirmed as a real physical effect, the morphological segregation
of clusters in the mass-\Ys\ plane would have major implications for
SZ surveys, which expect to produce nearly mass-limited cluster
samples minimally biased with respect to dynamical state.  However,
morphological segregation has not been seen in mass-\Ys\ relations
predicted from numerical simulations.  Indeed, it has generally been
found that the SZ signal is a robust mass proxy (merging and
non-merging clusters are co-located in the mass-\Ys\ plane) even
during periods of mass accretion \citep[e.g.,][]{MotlE05,PooleE07}.

Simulations of merging clusters do find that cluster mergers produce
transitory boosts in the SZ signal, although when integrated over the
entire cluster these boosts are small and short-lived \citep{PooleE07,
  WikE08}.  For example, \citeauthor{PooleE07} showed that for most
merger mass ratios and relative velocities, the observed $Y$ is
usually below the final ``post-merger'' value during the merger.  This
is a natural consequence of the finite time required to heat the ICM
to the higher equilibrated temperature of the post-merger halo.  The
observational signature of this effect is that at fixed post-merger
halo mass, merging (disturbed) clusters should have $Y$ suppressed
relative to non-merging (undisturbed) clusters. These theoretical
predictions are in the opposite sense to the morphological segregation
detected in our observed mass-\Ys\ relations.

\subsection{Correlation of Mass and $Y$ Measurements}
\label{sec:correlation}
The ideal scaling relation measurement compares two observables that
have been derived independently.  X-ray scaling relations have
typically been constructed from HSE-based masses derived from the same
X-ray data that appear in the observable axis.  HSE-based scaling
relations therefore inevitably suffer varying degrees of correlation
that depend on the details of the measurement methods and the
observable against which mass is plotted \citep[e.g.,][]{MantzE10-II}.
This correlation acts to suppress the scatter inferred for HSE-based
scaling relations.

The \Mwl\ and \Ys\ measurements presented in \S\ref{sec:analysis} are
also not completely independent, as the outer integration boundaries
for the SZ profile are set by the lensing-derived overdensity radii.
This introduces a correlation between observables.  The mass error
($\delta M \equiv M_{WL}-M_{true}$) for an individual cluster
corresponds to an error in the overdensity radius ($\delta r$),
\bq
\delta M = 4\pi r^2 \rho(r) \delta r~,
\label{eq:dM}
\eq
and this error is transferred to the calculation of \Ys.
Equation~\ref{eq:dM} can be converted to an equation involving
fractional errors in $M$ and $r$ by dividing by $M(r)=4\pi r^3
\bar\rho$/3,
\bq
\delta (log(M)) = 3\frac{\rho(r)}{\bar\rho}\delta (log(r))~.
\label{eq:dLM}
\eq
The perturbation in the calculated \Ys\ introduced by $\delta r$
follows the same form as equation~\ref{eq:dLM}, with the pressure
substituted for $\rho$ and \Ys\ for $M$.

From the equations for the logarithmic radial slopes of $M$ and \Ys\
we can infer the response of these two quantities to the errors in
$r_\Delta$.  The direction of motion in the \my\ plane for an error
$\delta r$, expressed as the slope $\delta log(M)/\delta log(Y)$, is
just the ratio of the logarithmic slopes in density ($\delta
log(M)/\delta log(r)$) and pressure at the radius of interest.  This
ratio varies from cluster to cluster and across overdensities, in the
present sample, it is (on average) 2.3 at \rtfh, but 1.7 at \rfh.  The
latter is very close to the slope of the self-similar scaling
relation, $1/B=5/3$.  On average then, the scatter between $M_{WL}$
and $M_{true}$ leads to errors in the derived \Ys\ that move the
clusters along the scaling relation at \rfh.

It was noted in \S\ref{s:scaleR} that the scatter between the mass
derived from the WL shear profile fitting technique and the true
cluster mass is expected to be $\sim$20\% \citep{BeckerKravtsov11},
which matches the values in Table~\ref{t:MY}.  That the observed and
predicted scatter agree so well may be coincidence; simulations
suggest that the astrophysics of the ICM can contribute another
10-15\% scatter between $Y$ and true mass, and systematic effects like
those discussed in subsequent sections can be expected to further
increase the observed scatter.  The above analysis suggests that
\SigYM\ could be larger if the \Ys\ and \Mwl\ measurements were
decoupled. The clusters in this sample all have \chandra\ data, which
provides us with an independent measurement of \rfh\ for the
calculation of \Ys.  Using the X-ray values of \rfh\ from
\citet{SandersonE09} as integration radii for both the NFW-halo fits
to the weak lensing data and the SZ profiles, we again find the 20\%
scatter in the scaling relation.  When these radii are used for the SZ
data alone, the scatter increases to 26\%.  The coupling between
observables through the integration radius therefore does appear to
lessen the observed scatter, and the scatter of fully independent
measurements may be more representative of the true intrinsic scatter.

\subsection{Choice of Pressure Profile}\label{sec:syssz}
Reliable measurements of \Ys\ depend on the appropriate choice of
pressure profile shape in the ICM models.  Systematic errors in the
choice of pressure profile may propagate to systematic errors in \Ys\
measurements, and thus contribute to the observed scatter and
segregation.  Our \Ys\ measurements employ the \citet{ArnaudE10}
average pressure profile. However, these authors noted that CC
clusters have more concentrated pressure profiles than disturbed
clusters\footnote{This cool-core/disturbed classification identifies
  ``disturbed'' clusters on the basis of \ww, and CC clusters on the
  basis of their central gas density \citep{PrattE09}.}.

As a test of the importance of the profile choice on our SZ
measurements, we recalculate our \Ys\ values using the
morphology-specific pressure profiles presented in \citet{ArnaudE10}.
For clusters we classify as ``disturbed'' and ``undisturbed'' we use
the \citeauthor{ArnaudE10} ``morphologically disturbed'' and
``cool-core'' profiles, respectively. The change in profile leads to
small systematic changes, the disturbed clusters decrease in \Ys\ by
1\% and 3\% at \rtfh\ and \rfh, respectively, and undisturbed clusters
increase by 1\% and 5\% at these radii.  These changes can, at most, reduce the
morphological segregation in the $\Mfh-\Ys$ relation from $41\%$ to
$31\%$. 

\begin{figure}
  \epsscale{1.15}
  \plotone{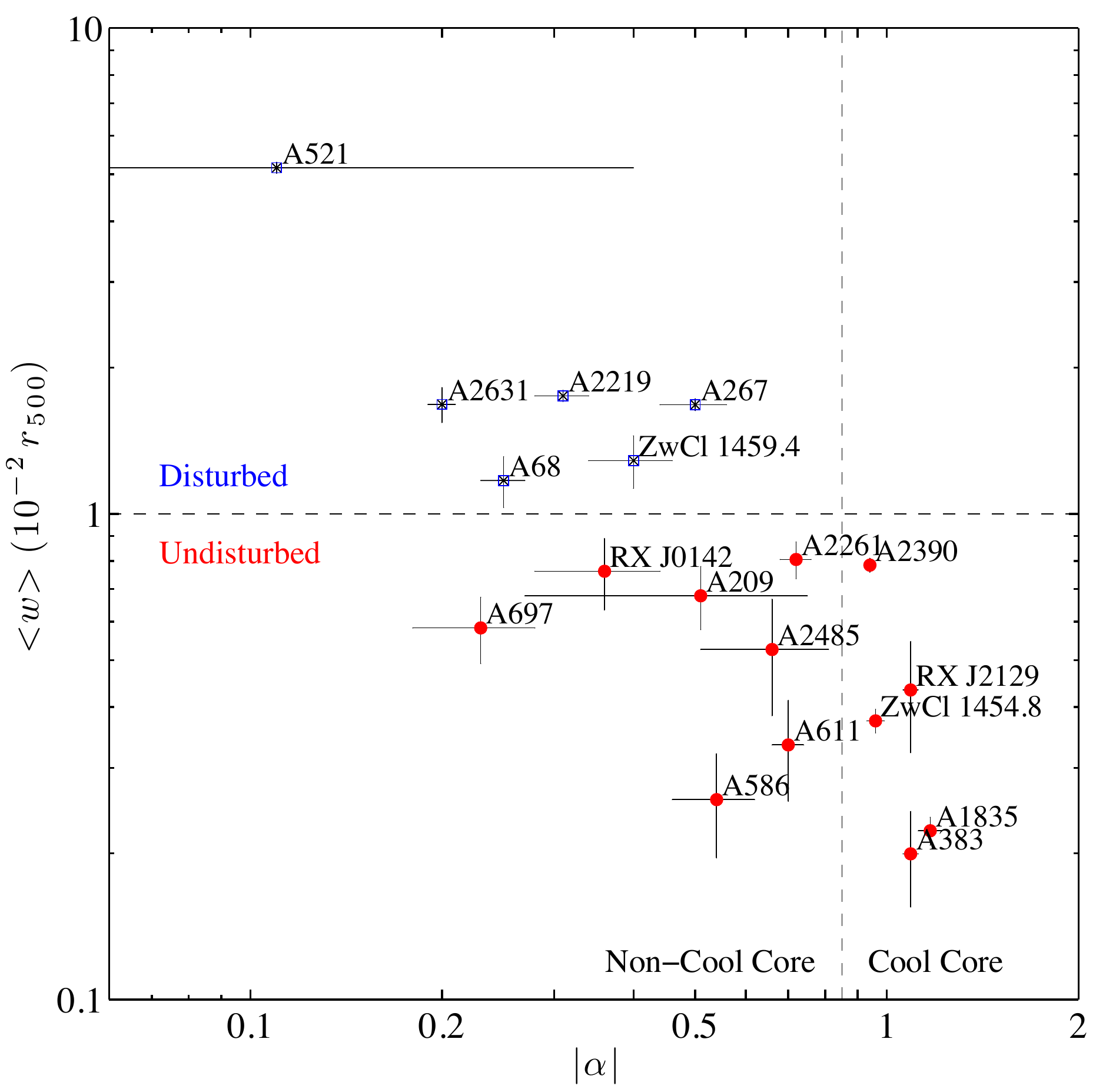}
  \caption{X-ray centroid shift versus slope of the gas density
    profile at $0.04\rfh$ ($|\alpha|$). The density slopes are taken
    from \citet{SandersonE09}, where $\alpha<-0.85$ provided a
    reliable dividing line between cool-core and non-cool core
    clusters.}
  \label{f:walpha}
\end{figure}

We expect this decrease in morphological segregation to be an upper
limit on the effect of pressure profile choice on our results.
Although we use the \citet{ArnaudE10} pressure profile for all of our
undisturbed clusters, not all of them host a cool core
(Figure~\ref{f:walpha}).  In the original measurement of the CC
profile, \citeauthor{ArnaudE10} excluded undisturbed clusters that
lacked cool cores from their subsample, and these excluded clusters
had profiles closest to the one we adopted in \S\ref{sec:szmodel}.  We
therefore conclude that although up to one fourth of the morphological
segregation can be attributed to systematic errors in \Ys\ due to
incorrect pressure profile choice in the ICM models, the majority of
the segregation remains to be explained.

\subsection{Choice of Weak-lensing Mass Model}\label{sec:syslensing}
Reliable measurements of $M_{\rm WL}$ also depend on appropriate model
choice.  The weak-lensing masses used in this article are based on the
initial \citetalias{OkabeE10} analysis of data from our Subaru
observing program, in which the mass distribution of each cluster was
described by a model containing a single spherical NFW halo
(\S\ref{sec:wlAnalysis}).  We therefore explore whether this choice
may contribute to the morphological segregation in the \my\ plane.

Additional halos that might be justified by the data include sub-halos
within each cluster, and large-scale structure projected along the
same line-of-sight.  We attempt the lowest order correction to our
lensing analysis by adding a second dark matter halo to each mass
model at the position of the most significant sub-peak in each of the
density maps presented in Appendix~1 of \citetalias{OkabeE10}.  These
2-halo models were fitted to the 2-dimensional weak-shear field using
{\sc lenstool}\footnote{\url{http://www.oamp.fr/cosmology/lenstool/}}
\citep{JulloE07}.  Comparing these models to the original 1-halo
models, we found that the Bayesian evidence ratio significantly
favored a two-halo model for all but 3 clusters.  We re-calculated
$M_{\rm WL}$ at $\Delta=500$ for each cluster using the more probable
of the two models, excluding the sub-halo if it is known to lie
outside the cluster (e.g.,\ RXJ\,0153.2$+$0102 adjacent to A\,267; see
Figure~21 in \citetalias{OkabeE10}), and assuming that the projected
separation of the two halos is the true three-dimensional separation
of the halos in all other cases.  In the resulting \Mfh-\Ys\ scaling
relation we find that neither the scatter nor segregation are reduced
within the uncertainties.

The one-dimensional radial profiles used to model the cluster
weak-lensing signal may also be a source of scatter.  \citet{OguriE10}
fitted elliptical weak-lensing mass models to \citetalias{OkabeE10}'s
data, thus relaxing the assumption of azimuthal symmetry in the single
NFW halo of the latter's models.  We find that the scatter and
segregation in the \my\ relation are all unchanged within the
uncertainties if \citeauthor{OguriE10}'s masses are substituted for
\citetalias{OkabeE10}'s masses.  Both of these tests suggest that the
description of the projected mass distribution is not the dominant
source of scatter.

\begin{figure}
 \epsscale{1.2}
 \plotone{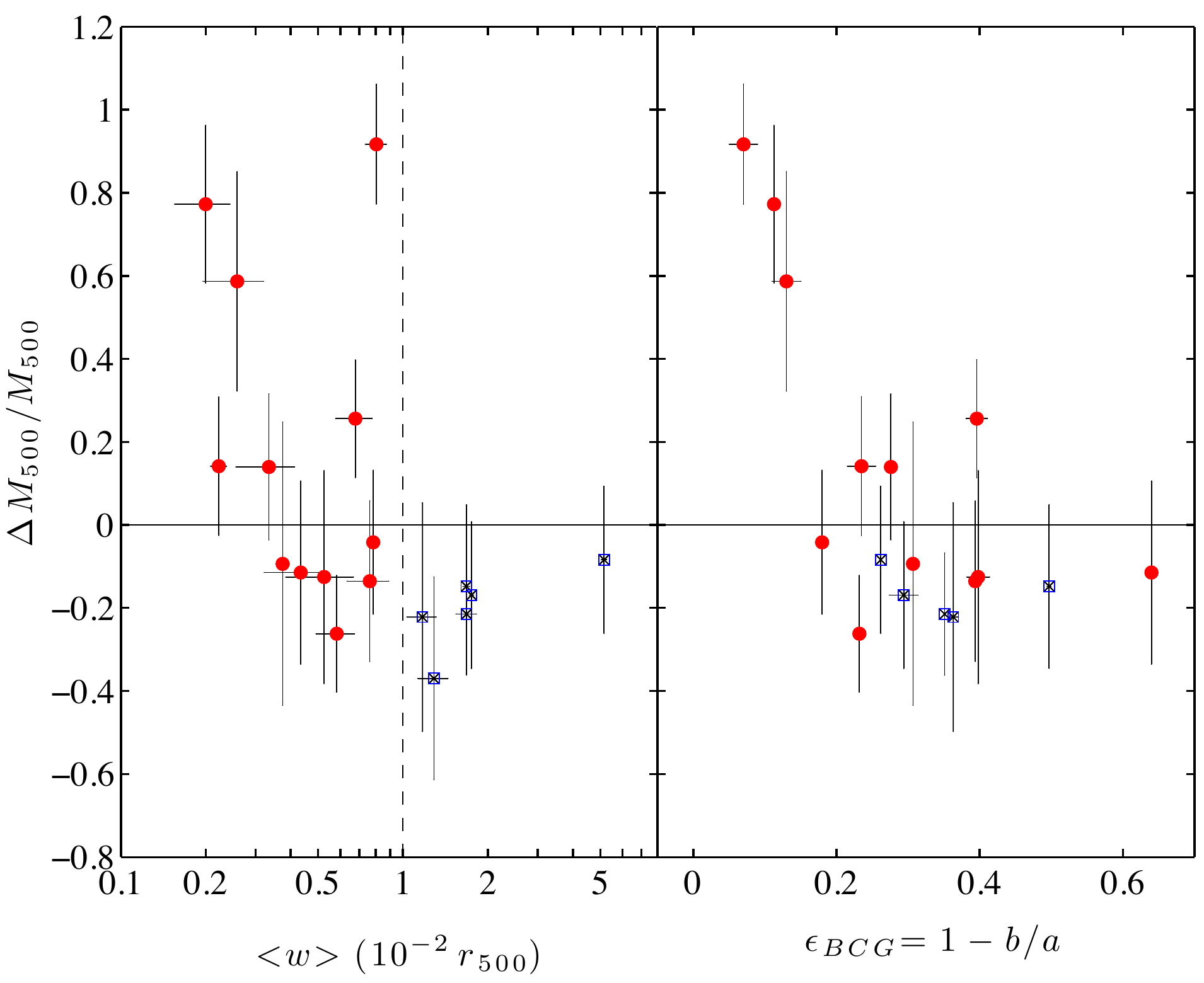}
 \caption{Fractional deviation in mass from the SS \Mfh-\Ys\ relation
   versus centroid shift \ww\ (left) and ellipticity of the BCG in
   each cluster (right).  Filled circles (red) and open squares (blue)
   represent undisturbed and disturbed clusters, respectively. The
   classification is done according to the centroid shift value and
   the dividing line is marked by the vertical dashed line in the left
   panel.}
 \label{f:devawe}
\end{figure}

The line-of-sight structure of the cluster dark matter is predicted to
introduce systematic errors to lensing masses derived assuming a
spherically symmetric profile
\citep[e.g.,][]{CorlessKing07,MeneghettiE10}.  An observational
indicator of the line-of-sight elongation of the mass distribution
would therefore be a powerful tool.  First, we consider the X-ray
centroid shift measurements upon which the disturbed/undisturbed
classification was made in \S\ref{sec:morph}.  We plot centroid shift
versus fractional deviation in mass from the best-fit SS scaling
relation in Figure~\ref{f:devawe}, finding no reliable trend between
the mass deviation and \ww.  Though the large-\ww\ clusters (disturbed)
lie exclusively on the low-mass side of the relation, as noted in
\S\ref{sec:scatter}, the small-\ww\ (undisturbed) clusters span a wide
range in deviation ($\Delta\Mfh/\Mfh\sim0-0.75$) at all centroid shift
values.  The largest deviations suggest that such clusters may be
prolate spheroids viewed along an axis close to their major-axis
\citep{CorlessKing07}, while the smallest deviations suggest that the
dark matter in such clusters is relatively undisturbed and spherical.
This implies that the interpretation of small centroid shifts is
degenerate between a relatively undisturbed ICM and the disturbed ICM
of a prolate cluster whose major axis is closely aligned with the
line-of-sight.  A ``cleaner'' indicator of the orientation of cluster
dark matter halos is therefore required.

BCGs are prolate stellar systems with their major axis closely aligned
with the major axis of the cluster dark matter halo and ICM
\citep[e.g.,][]{HashimotoE08, FasanoE10}.  The major axis of a prolate
BCG that is measured to be strongly elliptical in projection on the
sky likely lies close to the plane of the sky, and thus so too does
the major axis of the cluster dark matter halo.  An almost circular
BCG, on the other hand, indicates that the major axis is close to the
line-of-sight through the cluster.  BCG ellipticity measurements are
available from surface photometry of \emph{Hubble Space Telescope}
(\emph{HST}) observations of 15 of the 18 clusters in our sample
\citep{SmithE10}.  We analyzed two of the remaining clusters
(RX\,J0142.0$+$2131 and A\,697) in the same manner (the last cluster,
ZwCl\,1459.4$+$4240, has not been observed with \emph{HST}).  We find
that the clusters with the roundest BCGs suffer the strongest positive
deviations from the \my\ relation, and that clusters with the most
elliptical BCGs suffer negative deviations (right panel of
Figure~\ref{f:devawe}).  Most significantly, clusters with
$\epsilon_{\rm BCG}<0.15$ have exclusively large deviations
($\Delta\Mfh/\Mfh\gs0.5$).  The mean deviation of these ``round'' BCGs
($\epsilon<0.15$) is $\langle\Delta\Mfh/\Mfh\rangle=0.76\pm0.17$ and
of ``elliptical''BCGs ($\epsilon>0.15$) is
$\langle\Delta\Mfh/\Mfh\rangle=-0.10\pm0.17$, where the error bars are
the $1\sigma$ standard deviations on the means.

The correlation between $\Delta\Mfh/\Mfh$ and $\epsilon_\mathrm{BCG}$ is
strikingly similar to the trend between
$M_\mathrm{3D,WL}/M_\mathrm{3D,true}$ and viewing angle in the
numerically simulated clusters presented in \citet{MeneghettiE10}.
Indeed, visual inspection of \citeauthor{MeneghettiE10}'s Figure~17
suggests that we are viewing clusters with ``round'' BCGs within
$\sim20-30^\circ$ of the major axis of the cluster inertia ellipsoid,
and with ``elliptical'' BCGs at larger viewing angles.  An observed
BCG ellipticity of $\epsilon_\mathrm{BCG}=0.15$ (corresponding to an
observed axis ratio of $q=b/a=0.85$) can be converted to a viewing
angle with respect to the major axis by assuming an intrinsic BCG axis
ratio ($\beta$). We adopt $\beta=0.67$, following the measurements of
\citet{FasanoE10}.  The viewing angle is then estimated to be:
\begin{equation}
\psi=\arccos\left(\!\sqrt{\frac{1-(\beta/q)^2}{1-\beta^2}}\,\,\right)\simeq34^\circ
\end{equation}
which is consistent with the visual comparison of our
Figure~\ref{f:devawe} and Figure~17 of \citet{MeneghettiE10}.

Additional anecdotal support for this picture comes from detailed measurements of the 
three dimensional structure of Abell~383, which is the cluster with the second largest
value of $\Delta\Mfh/\Mfh$ in our sample and the outlier with the greatest effect on our
scaling relation fits (the lowest-\Ys\ cluster in Figure~\ref{f:MY}). \citet{NewmanE11} combined 
strong and weak lensing data, stellar kinematics in the BCG, and X-ray data to infer the three-dimensional 
shape of the cluster dark matter halo. They found that the cluster is very elongated along the line of sight, 
with an axis ratio of 2:1 between the line of sight and plane of sky. This agrees well with 
our inference that halo orientation is significantly affecting our weak lensing masses. 

\subsection{Implications for Scaling Relation Calibration}
Our data provide observational support for previous numerical studies
that have highlighted projection effects as an important source of
systematic uncertainties in weak-lensing cluster mass measurements.
These uncertainties arise from both the underlying shape of the main
cluster halo and sub-halos that reside within the cluster due to a
merger and/or in the surrounding large-scale structure.  Future
progress on the use of weak-lensing to calibrate mass-observable
scaling relations will therefore benefit from intrinsically asymmetric
lens models that include halo triaxiality and/or multiple halos.
Direct measurements of the ratio of the line-of-sight depth to the
plane-of-sky size of the cluster through joint X-ray and SZ
measurements of the ICM may aid in constructing a more
three-dimensional halo model. These efforts should lead to a reduction
in the scatter and segregation in the $M_{\rm WL}-\Ys$ relation.

While the mass scatter observed here is plausibly consistent with that predicted by lensing 
simulations, correlated scatter between mass and \Ys\ due to halo orientation and nearby structure 
will affect the scatter measured in scaling analyses like ours. Both the SZ and lensing signals can
be expected to suffer projection effects operating in the same direction, which will reduce
the scatter observed between \Ys\ and \Mwl\ from that predicted for $M_{\rm WL}-M_{\rm true}$ in
lensing simulations. The SZ projection effects differ from the lensing projection in important ways, 
which prevents perfect cancellation of the scatter. 
First, the ICM (in equilibrium) will follow isopotential contours 
rather than isodensity contours \citep{BuoteCanizares92}, which makes it much rounder than the dark matter.
\citet{LeeSuto03} show that the expected ICM eccentricity is typically 0.6 times that of the dark matter, 
corresponding to an ICM axis ratio of 1.2:1 for a matter axis ratio of 2:1. \citet{LauE11} show that for simulated clusters
the ICM is typically more spherical than either the dark matter or the potential at \rfh. Second, 
the SZ signal also depends on the ICM temperature, so that nearby but as-of-yet unassimilated 
halos will have a lower SZ signal per unit mass than the same material once it has passed through the virial shock of
the main halo. For this reason, projection of adjacent halos is a weaker effect for the SZ signal. 
Determinations of the scatter between \Ys\ and mass, as well as attempts at normalizing the \Ys-mass scaling relation
using weak lensing, will need to constrain the covariance of the two observables \citep[e.g.,][]{AllenEvrardMantz11}.

\section{Conclusions}
\label{sec:conclusions}

Surveys of galaxy clusters using the SZ effect require a
well-calibrated mass-observable scaling relation in order to constrain
cosmological parameters.  In this paper we report the first comparison
of the SZ observable \Ys\ with the weak lensing-determined halo mass.
We find a strong correlation, and our best-fit scaling relation is
consistent with expectations from self-similarity arguments
(Table~\ref{t:MY}).  Comparing to previous scaling relations measured
against hydrostatic masses, our normalization at \rtfh\ is consistent
with that of \citet{BonamenteE08} but at \rfh\ it mildly differs from
SPT results presented by \citet{AnderssonE11}. The scatter in our
relation is larger than predicted in simulations of the SZ signal,
though it is not unexpected given the difficulties inherent in
deriving spherical masses from the natively two-dimensional weak
lensing measurements.

We considered the morphology dependence of the \my\ relation by
dividing our sample into undisturbed and disturbed sub-samples based
on the well-established centroid shift parameter.  We found that
clusters segregate in the \my\ plane: the mass of undisturbed clusters
exceeds that of disturbed clusters at fixed \Ys. This effect becomes
more pronounced at larger radii (see Table~\ref{t:MY}), at
$\Delta=500$ undisturbed clusters are 30 -- 40\% more massive than disturbed
clusters.  Such a segregation is not predicted by numerical
simulations for comparisons of the SZ signal and true cluster mass.
We discuss a wide-range of possible interpretations of the
segregation, and conclude that the relative simplicity of the models
fitted to the data is the most likely cause.  No more than one fourth
of the segregation can be attributed to morphological biases
introduced by the adoption of a single (morphologically blind)
pressure profile in the SZ data analysis.  We infer that much of the segregation is caused by
modeling intrinsically prolate cluster dark matter halos as spherical
objects when determining weak lensing masses from our data.  
This assertion is supported by the similarity between our
measurements and predictions of projection-induced scatter for weak 
lensing data of similar quality, and the correlation we observe between 
BCG ellipticity, a proxy for halo orientation, and offset from our mean scaling relation.
Future uses of weak lensing as a calibration tool will therefore likely be aided by attempts
to correct for halo orientation, and we will explore this in the analysis of a larger
sample of objects in a future paper.

\acknowledgements
We thank Daisuke Nagai, Laurie Shaw, Neelima Sehgal, and Rebecca
Stanek for providing their simulated scaling relations.  We are
grateful to Gus Evrard, Andrey Kravtsov, Matthew Becker, and Bradford
Benson for comments that have improved this manuscript.  Support for
CARMA construction was derived from the Gordon and Betty Moore
Foundation, the Kenneth T. and Eileen L. Norris Foundation, the James
S. McDonnell Foundation, the Associates of the California Institute of
Technology, the University of Chicago, the states of California,
Illinois, and Maryland, and the National Science Foundation. Ongoing
CARMA development and operations are supported by the National Science
Foundation under a cooperative agreement, including grant AST-0838187
at the University of Chicago, and by the CARMA partner universities.
Partial support is provided by NSF Physics Frontier Center grant
PHY-0114422 to the Kavli Institute of Cosmological Physics.  D. P. M.
was supported for part of this work by NASA through Hubble Fellowship
grant HST-HF-51259.01. He acknowledges the Kavli Institute for
Theoretical Physics for its hospitality during part of this research,
supported by NSF grant PHY-0551164.  G. P. S. acknowledges support
from the Royal Society.  Support for T. M. was provided by NASA
through the Einstein Fellowship Program, grant PF0-110077.  N. O. was
partially supported by a Grant-in-Aid (0740099), and this work was
supported by the programs ``Weaving Science Web beyond Particle-Matter
Hierarchy'' in Tohoku University and ``Probing the Dark Energy through
an Extremely Wide and Deep Survey with Subaru Telescope'' (18072001),
all of which are funded by the Ministry of Education, Culture, Sports,
Science, and Technology of Japan.  Y.Y.Z. acknowledges support from
the German BMBF through the Verbundforschung under grant
No.\,50\,OR\,1005.

{\it Facilities:} \facility{CARMA}

\end{document}